\documentclass[sigconf,table]{acmart}
\usepackage[frozencache,cachedir=.]{minted}
\usepackage{comment}
\usepackage{algorithmic}
\usepackage{graphicx}
\usepackage{textcomp}
\usepackage[]{xcolor} 
\usepackage{hyperref}
\usepackage{multirow,booktabs}
\usepackage{heatmap} 
\usepackage{enumitem}           
\usepackage[caption=false]{subfig} 

\usepackage[most]{tcolorbox}
\usepackage[parfill]{parskip}
\tcbset{
  parskip/.style={
    before={\par\pagebreak[0]\parindent=0pt},
    after={\parfillskip=0pt\par},
  },
}
\newtcolorbox{resultbox}[2][]{%
    colback=black!5,
    colframe=black!5,
    notitle,
    sharp corners,
    borderline west={2pt}{0pt}{red!80!black},
    enhanced,
    breakable,
    boxsep=0pt,
    left=8pt,right=2pt,top=2pt,bottom=2pt,
    }

\newcommand{\ie}{\textit{i.e.},\ }
\newcommand{\eg}{\textit{e.g.},\ }
\newcommand{\etal}{\textit{et al.} }
\newcommand{\etc}{{\em etc.}}

\newcommand{\finding}[1]{#1} 

\newcommand{\figref}[1]{Fig.~\ref{#1}}
\newcommand{\secref}[1]{Section~\ref{#1}}
\newcommand{\listref}[1]{Listing~\ref{#1}}
\newcommand{\tblref}[1]{Table~\ref{#1}}

\newcommand*\circled[1]{\tikz[baseline=(char.base)]{
            \node[thick=1.5pt,shape=circle,draw,inner sep=0.5pt] (char) {\small\textbf{#1}};}}

\definecolor{codebg}{rgb}{0.99,0.99,0.99}
\definecolor{hiliteColor}{rgb}{1,0.92549019607,0.6}
\definecolor{tainted}{rgb}{0,1,1}

\newmintinline{python}{fontsize=\scriptsize}
\newminted[PythonSourceCode]{python}{
    labelposition=topline,
    frame=single,
    numbersep=2pt,
    fontsize=\scriptsize,
    xleftmargin=5pt,
    framerule=0.5pt,
    linenos, rulecolor=\color{gray!60}
}

\newmintinline{java}{fontsize=\scriptsize}
\newminted[JavaSourceCode]{java}{
    labelposition=topline,
    frame=single,
    numbersep=2pt,
    fontsize=\scriptsize,
    xleftmargin=5pt,
    framerule=0.5pt,
    linenos, rulecolor=\color{gray!60}
}

\definecolor{magnolia}{rgb}{0.97, 0.96, 1.0}
\newmintinline[code]{python}{fontsize=\small}
\newmintinline[codePerl]{perl}{fontsize=\small,bgcolor=magnolia}
\newmintinline[codeJava]{java}{fontsize=\small}

\usepackage{tikz}
\usetikzlibrary{calc,tikzmark}

\colorlet{colorMain}{green!10}
\colorlet{colorSum}{yellow!20}

\newcommand{\nEvosuiteProjs}[0]{47}  
\newcommand{\nJavaHumanEval}[0]{160}  
\newcommand{\nEvosuiteCUTs}[0]{194}  
\newcommand{\nEvosuiteMUTs}[0]{411}  

\newcommand{\StarCoder}[0]{{StarCoder}}
\newcommand{\gpt}[0]{{GPT-3.5-Turbo}}
\newcommand{\codexTwo}[0]{{Codex (2K)}}
\newcommand{\codexFour}[0]{{Codex (4K)}}

\AtBeginDocument{%
  }

\acmConference[EASE 2024]{The 28th International Conference on Evaluation and Assessment in Software Engineering}{18–21 June, 2024}{Salerno, Italy}
\setcopyright{acmcopyright}
\copyrightyear{2024}
\acmYear{2024}
\acmDOI{XXXXXXX.XXXXXXX}

\begin{document}

\title{Using Large Language Models to Generate JUnit Tests: An Empirical Study}


\author{Mohammed Latif Siddiq}
\email{msiddiq3@nd.edu}
\orcid{0000-0002-7984-3611}
\affiliation{%
  \institution{University of Notre Dame}
  \city{Notre Dame}
  \state{IN}
  \country{USA}
  \postcode{46556}
}

\author{Joanna C. S. Santos}
\email{joannacss@nd.edu}
\orcid{0000-0001-8743-2516}
\affiliation{%
  \institution{University of Notre Dame}
  \city{Notre Dame}
  \state{IN}
  \country{USA}
  \postcode{46556}
}

\author{Ridwanul Hasan Tanvir}
\email{rpt5409@psu.edu}
\affiliation{%
  \institution{Pennsylvania State University}
  \city{University Park}
  \state{PA}
  \country{USA}
  \postcode{16802}
}
\author{Noshin Ulfat}
\email{noshin.ulfat@iqvia.com}
\affiliation{%
  \institution{IQVIA Inc.}
  \city{Dhaka}
    \country{Bangladesh}
}

\author{Fahmid Al Rifat}
\email{fahmid@cse.uiu.ac.bd}
\affiliation{%
  \institution{United International University}
  \city{Dhaka}
    \country{Bangladesh}
}
\author{Vinícius Carvalho Lopes}
\email{vlopes@nd.edu}
\affiliation{%
  \institution{University of Notre Dame}
  \city{Notre Dame}
  \state{IN}
  \country{USA}
  \postcode{46556}
}

\begin{abstract}
A code generation model generates code by taking a prompt from a code comment, existing code, or a combination of both. 
Although code generation models (\eg GitHub Copilot) are increasingly being adopted in practice, it is unclear whether they can successfully be used for unit test generation without fine-tuning for a strongly typed language like Java. To fill this gap, we investigated how well three models (Codex, GPT-3.5-Turbo, and StarCoder) can generate unit tests. We used two benchmarks (HumanEval and Evosuite SF110) to investigate the effect of context generation on the unit test generation process. We evaluated the models based on compilation rates, test correctness, test coverage, and test smells. We found that the Codex model achieved above 80\% coverage for the HumanEval dataset, but no model had more than 2\% coverage for the EvoSuite SF110 benchmark. The generated tests also suffered from test smells, such as Duplicated Asserts and Empty Tests.
\end{abstract}



\maketitle

\section{Introduction}
\textit{\textbf{Unit testing}}~\cite{beck2003TDD} is a software engineering activity in which individual units of code are tested in isolation.
This is an important activity because it helps developers identify and fix defects early on in the development process and understand how the various units of code in a software system fit together and work as a cohesive whole. 
Despite its importance, in practice, 
developers face difficulties when writing unit tests~\cite{runeson2006survey,kochhar2015understanding, thomas2002mock,daka2015Modeling}. This leads to a negative effect: developers may not write tests for their code. In fact, a prior study~\cite{gonzalez2017large} showed that out of 82,447  studied GitHub projects, only \textbf{17\%} of them contained test files.

Since implementing test cases to achieve good code coverage is a time-consuming and error-prone task,  prior works \cite{sina18unittest,tufano@2020} developed techniques to automatically generate unit tests. Although automatically generated unit tests help increase code coverage~\cite{serra2019effectiveness,bacchelli2008effectiveness}, they are still not frequently used in practice~\cite{fraser2015does}.

With the advances of large language models (LLMs), LLM-based code generation tools (\eg  GitHub Copilot) are increasingly becoming part of day-to-day software development. A survey of 500 US-based developers showed that 92\% of them are using LLM-based coding assistants both for work and personal use~\cite{shani2023survey}.  Part of this fast widespread adoption is that LLMs automate repetitive tasks so that they can focus on higher-level, challenging tasks~\cite{albert22}. 
With the increasing popularity of code generation LLMs, prior works investigated the correctness of the generated code~\cite{dakhel2022github}, their quality~\cite{siddiq2022empirical}, security~\cite{pearce2022asleep} and whether they can be used for API learning tasks \cite{apiicpc22}, and code complexity prediction \cite{siddiq23zero}. However, it is currently unclear the effectiveness of using prompt-based pre-trained code generation models to generate \textit{unit tests} for strongly typed languages such as Java. In fact, prior works \cite{cassano2023multipl,mbxp_athiwaratkun2022} have shown that LLMs perform better for weakly typed languages (\eg Python and JavaScript) but not as well for strongly typed languages. This is partially due to the limited training sets availability and the fact that strongly typed languages have strict type-checking that can prevent the code from even compiling. 

In light of this research gap,
we conducted an empirical study using three LLMs (Codex~\cite{chen2021codex},  GPT-3.5-Turbo~\cite{chatgpt} and StarCoder \cite{li2023starcoder}) to generate JUnit5 tests for  classes in the HumanEval dataset's Java version~\cite{2023multilingual} and \nEvosuiteProjs{} open-source projects from the SF110 dataset~\cite{evosuite}. In our study, we investigate how well LLMs can generate JUnit tests (RQ1) and how different context styles (\eg only using the method under test, the presence/ absence of JavaDocs \etc) provided as input to an LLM can influence its performance (RQ2). We examined the generated tests with respect to their \textit{compilation rates}, \textit{correctness}, \textit{code coverage}, and quality (in terms of test smells). 
While concurrent works \cite{lemieux@2023, schafer@2023} studied the usefulness of LLM as a \textit{helper} for search-based test generation techniques on weakly typed languages  (\ie Python and JavaScript),    our work investigates whether LLMs can be used off-the-shelf to generated unit tests for a strongly-typed language like Java. Moreover, we examine these generated tests in terms not only of their \textit{correctness}, but also their \textit{quality}, as well as the effectiveness of different \textit{context styles}.

The \textbf{contributions} of our work are:
\circled{1} A systematic study of three LLMs for zero-shot unit test generation for \nEvosuiteCUTs{} classes from \nEvosuiteProjs{} open-source projects in the SF110 dataset~\cite{SF100} and \nJavaHumanEval{} classes from the HumanEval dataset~\cite{2023multilingual}. 
\circled{2} An investigation of the quality of the produced unit tests by studying the prevalence of test smells in the generated unit tests.
\circled{3} A comparison of how different context styles affect the performance of LLMs in generating tests.
\circled{4} A  discussion about the implication of using code generation models for unit test generation in a Test Driven Development (TDD) environment. 
All the scripts used to gather the data and spreadsheets compiling all the results are available on Zenodo~\footnote{\url{https://doi.org/10.5281/zenodo.10530787}}.

\section{Background}\label{sec:background}

\subsection{Unit Tests \& Test Smells}
The goal of \textit{\textbf{unit testing}} is to validate that each program unit is working as intended and meets its requirements~\cite{tim1999}. 
A \textit{\textbf{unit}} refers to a piece of code that can be isolated and examined independently (\eg functions/methods, classes, \etc). Just like production code,  unit tests need to be not only \textit{correct} but also satisfy other quality attributes, such as \textit{maintainability} and \textit{readability}~\cite{gonzalez2017large}. 

\textit{\textbf{Unit test smells}} (henceforth ``test smells'') are indicators of potential problems, inefficiencies, or bad programming/design practices in a unit test suite \cite{guerra7,Michaela13,fabio16,Simula.SE.525,peruma19}. 
There are many test smell types, ranging from tests that are too slow/fragile to tests that are too complex or too tightly coupled to implementing the code under test~\cite{meszaros03}.
For example, the Java code in \listref{lst:exampleunittestsmell} has a unit test for a method from the \codeJava{LargestDivisor} class. 
It checks whether the Method Under Test (MUT) returns the largest divisor of a number. Although this test is correct, there is no explanation for the expected outputs passed to the assertions, which is a case of the \textsf{Magic Number Test} smell~\cite{meszaros03}. It also has multiple assertions in the same test method, an example of \textsf{Assertion Roulette} smell~\cite{Simula.SE.525}.

\vspace{-5pt}

\begin{listing}[H]
{
\begin{JavaSourceCode*}{label=\textcolor{black}{\tiny{LargestDivisorTest.java}}}
public class LargestDivisorTest {
    @Test
    void testLargestDivisor() {
        assertEquals(5, LargestDivisor.largestDivisor(15));
        assertEquals(1, LargestDivisor.largestDivisor(3));
    }
}
\end{JavaSourceCode*}
}\vspace{-10pt}
\caption{Example of Unit Test and Unit Test Smell}\label{lst:exampleunittestsmell}
\end{listing}

\vspace{-10pt}


\subsection{Code Generation}

\textbf{\textit{Large Language Models}} (LLMs) are advanced AI systems capable of understanding and generating human-like text. They can bed used to answer questions, create content, and even engage in conversation. \textbf{\textit{Code LLMs}} (henceforth simply ``LLMs'') are a specialized type of LLMs trained  on source code to help with code-related tasks, \eg code completion \cite{izadi2022codefill,kim2021code,svyatkovskiy2021fast}, search \cite{codebert}, and summarization~\cite{gao2022m2ts}. They are designed to generate source code from a given \textit{\textbf{prompt}}~\cite{allamanis2018survey}, such as a text written in natural language, pseudocode, code comments \etc~These techniques may also take into  account  the surrounding \textit{\textbf{context}} when generating the code, such as file/variable names, other files in the software system, \etc~




\section{Methodology}\label{sec:method}

In this work, we answer two research questions.


\begin{itemize}[leftmargin=21pt,noitemsep,topsep=0pt,itemsep=0pt]
    \item[\textbf{RQ1}] \textit{\textbf{How well can LLMs  generate JUnit tests?}}
\end{itemize}
We used  \gpt{}, StarCoder, and Codex to generate unit tests for  competitive programming assignments from the Java version of the HumanEval dataset~\cite{2023multilingual} 
as well as  \nEvosuiteProjs{} open-source projects from the EvoSuite SF110 benchmark dataset\cite{evosuite}.
We measured the LLMs' performance by computing the test's branch/line coverage, correctness, and quality (in terms of test smells). We also compared the performance of these models with Evosuite~\cite{evosuite}, an existing state-of-the-art approach.


\begin{itemize}[leftmargin=21pt,noitemsep,topsep=0pt,itemsep=0pt]
    \item[\textbf{RQ2}] \textit{\textbf{How do different code elements in a context influence the performance of LLMs in generating JUnit tests?}}
\end{itemize}\vspace{-2pt}
When developers use LLMs to generate JUnit tests, they create a \textit{\textbf{prompt}} (\eg \textit{``Write a JUnit test to verify that \codeJava{login(req)} returns ...''}) and the method (unit) under test becomes the \textit{\textbf{context}} for that prompt. Since the unit under test (context) can include several \textit{code elements}, we  investigate how these different elements affect the generated tests. 
To answer RQ2, we conducted a controlled experiment in which we created 3 different scenarios for the  HumanEval~\cite{chen2021codex,2023multilingual}, and 4 scenarios for \nEvosuiteProjs{} open-source projects from the EvoSuite SF110 dataset\cite{evosuite}.
Each scenario contains a different set of code elements. Then, we use Codex, GPT-3.5-Turbo, and StarCoder to generate JUnit tests for each scenario. We measured their performance  in terms of compilation rates, code coverage, the total number of correct unit tests, and the incidence of test smells.

\subsection{Answering RQ1}\label{sec:RQ1}

We followed a three-step systematic process to investigate how well LLMs can generate unit tests:
\circled{1} we collected \textbf{\nJavaHumanEval} Java classes from the \textbf{multilingual HumanEval dataset}~\cite{2023multilingual} and \textbf{\nEvosuiteCUTs{}} Java classes from \textbf{\nEvosuiteProjs} projects in the \textbf{Evosuite SF110 benchmark dataset}~\cite{SF100,campos2014continuous}; \circled{2} we generated JUnit5 tests using three LLMs; \circled{3} we computed the compilation rates, correctness, number of smells, as well as the line/branch coverage for the generated tests and compared with Evosuite v1.2.0, which is a state-of-the-art unit test generation tool~\cite{evosuite}. In this paper, \textbf{\textit{methods}} are our units under test. 

\subsubsection{Data Collection}

We use the \textsf{multilingual HumanEval dataset}~\cite{2023multilingual} because it has been widely used in prior works~\cite{Nijkamp2022ACP, incoder, siddiq2022empirical} to evaluate code LLMs. Similarly, we use the \textsf{SF110} dataset because it is a popular benchmark for unit test generation~\cite{Fraser14}. 

\begin{listing}[H]
\centering
\begin{JavaSourceCode*}{label=\textcolor{black}{\tiny{GreatestCommonDivisor.java}},highlightlines={1-7}}
class GreatestCommonDivisor {
    /**
     * Return the greatest common divisor of two integers a and b.
     * > greatestCommonDivisor(3, 5)
     * 1
     */
    public static int greatestCommonDivisor(int a, int b) {
        if (a == 0) return b;
        return greatestCommonDivisor(b % a, a);
    }
}
\end{JavaSourceCode*}
\vspace{-10pt}\caption{Sample from the extended HumanEval~\cite{2023multilingual}}
\label{lst:truncatenumber}
\end{listing}\vspace{-12pt}

\noindent-- The \textbf{multilingual HumanEval dataset}~\cite{2023multilingual} contains \textbf{\nJavaHumanEval} \textit{prompts} describing programming problems for Java and other programming languages crafted from the original Python-based HumanEval~\cite{chen2021codex}. However, this multilingual version does not provide a solution for each prompt. Thus, we wrote the solution for each problem and tested our implementation using the provided test cases. 
\listref{lst:truncatenumber} shows a sample taken from this dataset, where the prompt is in lines 1--7 and the solution is in lines 8--11.

\noindent -- The \textbf{SF110 dataset}, which is an Evosuite benchmark consisting of 111 open-source Java projects
retrieved from SourceForge. This benchmark contains 23,886 classes, over 800,000 bytecode-level branches, and 6.6 million lines of code \cite{Fraser14}.

\paragraph{\underline{\textit{Class and Method Under Test Selection}}}


Each class in the multilingual HumanEval~\cite{2023multilingual} has one \codeJava{public static} method and may also contain private ``helper'' methods to aid the solution implementation. In this study, \textbf{\textit{all}} the public static methods are selected as  methods under test (\textbf{MUTs}). 

For the SF110 benchmark, 
we first retrieved only the classes that are \code{public} and \textbf{\textit{not}} \code{abstract}. We then discarded test classes (\ie placed on a \textsf{src/test} folder, or that contains the keyword ``\textsf{Test}'' in its name). Next, we identified \textit{testable methods} by applying  \textit{inclusion} and \textit{exclusion} criteria. The \textit{exclusion} criteria are applied to the \textbf{\textit{non-static}} methods that 
\textbf{(E1)} have a name starting with ``get'' and takes no parameters, \textbf{\textit{or}} \textbf{(E2)} have a name starting with ``is'', takes no parameter and returns a \codeJava{boolean} value, \textbf{\textit{or}} \textbf{(E3)} have a name starting with ``set'', \textbf{\textit{or}} \textbf{(E4)} override the ones from \code{java.lang.Object} (\ie \codeJava{toString()}, \codeJava{hashCode()}, \etc). The exclusion criteria \textbf{E1--E3} are meant to disregard ``getter'' and ``setter'' methods. The inclusion criteria are that the method has \textbf{(I1)} a \codeJava{public} visibility, \textbf{(I2)} a \codeJava{return} value, \textbf{\textit{and}} \textbf{(I3)} a \textit{good} JavaDoc. A \textit{good} JavaDoc is one that \textbf{(i)} has a description \textit{or} has a non-empty \codeJava{@return} tag, and \textbf{(ii)} all the method's parameters have an associated description with \codeJava{@param} tag. After this step, we obtained a total of  {{30,916}} methods under test (MUTs) from {{2,951}} classes.
Subsequently, we disregard projects based on the number of retrieved testable methods (MUTs). We kept projects with at least one testable method (\ie first quartile) and at most 31 testable methods (\ie third quartile), obtaining a total of {53} projects. This filtering aimed to remove projects with too \textit{little} or too \textit{many} MUTs, which would exceed the limit of the number of tokens that the models can generate.
We then removed {6} of these projects in which we could not compile their source code, obtaining \textbf{{\nEvosuiteProjs{}}} projects and a total of \textbf{\nEvosuiteMUTs{}} MUTs from \textbf{\nEvosuiteCUTs{}} classes.


\subsubsection{Unit Test Generation}\label{subsubsec:RQ1TestGen}

We used Codex, GPT-3.5-Turbo, and StarCoder to generate JUnit tests.
\textbf{Codex} is a 12 billion parameters LLM \cite{chen2021codex} descendant of the GPT-3 model \cite{brown20} which powers GitHub Copilot. In this study, we used \textsf{code-davinci-002}, the most powerful codex model version of Codex.
\textbf{GPT-3.5-turbo} is the model that powers the ChatGPT chatbot. It allows multi-turn conversation, and it can be instructed to generate code~\cite{chatgpt}. 
\textbf{StarCoder} is a 15.5 billion parameter open-source code generation model with 8,000 context length and has infilling capabilities. 
In this work, we used the base model from the StarCoder code LLM series. 

To generate the JUnit tests, we performed a two-step process:

\noindent\textbf{\circled{1} \textit{Context and Prompt Creation}}: We created a \textit{\textbf{unit test prompt}} (henceforth ``prompt''), which instructs the LLM to generate \textbf{10} test cases for a specific method, and a \textit{\textbf{context}}, which encompasses the whole code from the method's declaring class as well as import statements to core elements from the JUnit5 API. \listref{lst:prompt_java} illustrates the structure of a prompt and context, in which lines 1-9 and lines 10-20 are part of the \textit{context} and \textit{prompt}, respectively. 
The context starts with a comment indicating the class' file name followed by its full code (\ie its package declaration, imports, fields, methods, \etc). 
Similarly, the prompt starts with a comment indicating the expected file name of the generated unit test. 
Since a class can have more than one testable method, we generated one unit test file for each testable method  in a class and appended a suffix to avoid duplicated test file names. A suffix is a  number that starts from zero. After this code comment, the prompt includes the same package declaration and  import statements from the class. It also has import statements to the \code{@Test} annotation and the \texttt{\small assert*} methods (\eg \codeJava{assertTrue(...)})  from JUnit5. Subsequently, the prompt contains the test class' JavaDoc that specifies the MUT, and how many test cases to generate. The prompt ends with the test class declaration followed by a new line (\code{\n}), which will trigger the LLM  to generate code to complete the test class declaration.

\vspace{-5pt}
\begin{listing}[!ht]
{\renewcommand\theFancyVerbLine{%
\rmfamily\tiny\ifnum\value{FancyVerbLine}=25 
  \setcounter{FancyVerbLine}{25}\ldots
\else
\arabic{FancyVerbLine}%
\fi
}
\begin{JavaSourceCode*}{label=\textcolor{black}{\tiny{classNameSuffixTest.java}},highlightlines={1-9}}
// ${className}.java
${packageDeclaration}
${importedPackages}
class ${className}{
  /* ... code before the method under test  ... */
  public ${methodSignature}{ /* ... method implementation ... */ }
  /* ... code after the method under test  ... */
}

// ${className}${suffix}Test.java
${packageDeclaration}
${importedPackages}
import org.junit.jupiter.api.Test;
import static org.junit.jupiter.api.Assertions.*;

/**
* Test class of {@link ${className}}. 
* It contains ${numberTests} unit test cases for the 
* {@link ${className}#${methodSignature}} method.
*/
class ${className}${suffix}Test {
    
\end{JavaSourceCode*}
}
\vspace{-10pt}\caption{Prompt template for RQ1}\label{lst:prompt_java}
\end{listing}
\vspace{-5pt}

\noindent\textbf{\textit{\circled{2} Test Generation}}: 
Although all used LLMs  can generate code, they have technical differences in terms of number of tokens they can handle. Thus, we took slightly different steps to generate tests with these LLMs. 
We used the OpenAI API to generate tests using the \textbf{Codex} model. Codex can take up to 8,000 tokens as input and generate up to 4,000 tokens. Thus, we configured this model in two ways: one to generate up to 2,000 tokens and another to generate up to 4,000 tokens. We will call each of them \textbf{Codex (2K)} and \textbf{Codex (4K)}, respectively.  For both cases, we set the model's \texttt{temperature} as zero  in order to produce more deterministic and reproducible output motivated by previous studies \cite{9809175,chen2022codet, Savelka23}. The rest of its inference parameters are set to their default values. 

\textbf{GPT-3.5-Turbo} is also accessible via the OpenAI API. It can take up to 4,096 tokens as input and generate up to 2,048 tokens. We asked this LLM to generate up to 2,000 tokens and dedicated the rest (2,096) to be used as input. Its temperature is also set to zero and the other parameters are set to their defaults. Moreover,   we set the \textsf{system} role's  content to \textit{``You are a coding assistant. You generate only source code.''} and 
the \textsf{user} role's content to the context and prompt. Then, the \textsf{assistant} role outputs the generated test. 
For \textbf{StarCoder}, we used the \textsf{StarCoderBase} model available on HuggingFace library\footnote{\url{https://huggingface.co}}. It has an 8,000 tokens context window combining the input prompt tokens and the output tokens. We limit the output token to 2,000 tokens to align the experiment with the other two models. We also keep the same inference parameters as the Codex model.

\subsubsection{Data Analysis and Evaluation}\label{subsec:DataAnalysis}
We compiled all the unit tests  together with their respective production code and required libraries.
As we compiled the code and obtained compilation errors, we observed that several of these errors were caused by simple \textit{syntax} problems that could be automatically fixed through \textit{heuristics}. Specifically, we noticed that LLMs may
\textit{\textbf{(i)}} generate an \textit{extra} test class that is incomplete, 
\textbf{\textit{(ii)}} include natural language explanations before and/or after the code,
\textbf{\textit{(iii)}} repeat the class under test and/or the  prompt, 
\textbf{\textit{(iv)}} change the package declaration or \textbf{\textit{(v)}} remove the package declaration, 
\textbf{\textit{(vi)}} generate integer constants higher than \codeJava{Integer.MAX_VALUE}, 
\textbf{\textit{(vii)}} generate incomplete unit tests after it reaches its token size limit.  
Thus, we developed \textbf{7} heuristics (\textbf{H1}--\textbf{H7}) to automatically fix these errors
:

\begin{enumerate}[leftmargin=*,noitemsep,topsep=0pt,itemsep=0pt]
    \item[\scriptsize \textbf{H1}] It removes any code found \textit{after} any of the following patterns: \codePerl{"</code>"}, \codePerl{"\n\n// {CUT_classname}"}, and \codePerl{"\n```\n\n##"}. 
    \item[\scriptsize \textbf{H2}] It keeps code snippets within backticks (\ie \codePerl{``` code ```}) and removes any text before and after the backticks. 
    \item[\scriptsize \textbf{H3}] It removes the original  prompt from the generated unit test. 
    \item[\scriptsize \textbf{H4}] It finds the package declaration in the unit test and renames it to the package of the CUT.  
    \item[\scriptsize \textbf{H5}] It adds the package declaration if it is missing. 
    \item[\scriptsize{\textbf{H6}}] It replaces large integer constants by \codeJava{Integer.parseInt(n)}. 
    \item[\scriptsize{\textbf{H7}}] It fixes incomplete code by iteratively deleting lines (from bottom to top) and adding 1-2 curly brackets. At each iteration, it removes the last line and adds one curly bracket. If the syntax check fails,  it adds two curly brackets and checks the syntax again. If it fails, it proceeds to the next iteration by removing the next line (bottom to top). The heuristic stops if the syntax check passes or it finds the class declaration (\ie ``\codeJava{class ABC}''), whichever condition occurs first.  
        
\end{enumerate}

\paragraph{\underline{\textit{Metrics}}}

We ran each generated unit test  with JaCoCo~\cite{jacoco} to compute the \textbf{\textit{line coverage}}, \textbf{\textit{branch coverage}} and \textbf{\textit{test correctness}} metrics.
\textbf{\textit{Branch Coverage}}~\cite{codecoveragegoogle} measures how many branches are covered by a test, \ie $\frac{Number~of~visited~branches}{Total~number~of~branches} \times 100$.
\textbf{\textit{Line Coverage}} measures how many lines were executed by the unit test out of the total number of lines~\cite{michael18}, \ie $\frac{Number~of~executed~lines}{Total~number~of~lines} \times 100$.
\textbf{\textit{Test Correctness}} measures how effectively an LLM generates correct input/output pairs. We assume that the code under test is implemented correctly. The reasoning behind this assumption is twofold: the HumanEval dataset contains common problems with well-known correct solutions, and the SF110 projects are \textbf{\textit{mature}} open-source projects. Given this assumption, a failing test case is considered to be \textit{incorrect}. Thus, we compute the number of generated unit tests that did not fail.

We ran the tests using a timeout of \textbf{2} and \textbf{10} minutes for the HumanEval and the SF110 datasets, respectively, because we observed generated tests with infinite loops.  Moreover, we analyzed the quality of the unit test from the perspective of the \textbf{test smells}. To this end, we used \textsc{TsDetect}, a state-of-the-art tool that detects 20 test smell types~\cite{peruma2020tsdetect,peruma19}. Due to space constraints, we provide a list of the test smells detectable by \textsc{TsDetect} with their descriptions in our replication package.

\subsection{RQ2: Code Elements in a Context}


To investigate how different code elements in a context influence the generated unit test,  we first 
created \textbf{\textit{three}} scenarios for the HumanEval dataset and \textbf{\textit{four}} for the Evosuite Benchmark. 


\noindent\textbf{\underline{HumanEval Scenarios}}: Recall that each MUT in this dataset has a JavaDoc describing the method's expected behavior and examples of input-output pairs (see \listref{lst:exampleunittestsmell}). Thus, we created one scenario (\textbf{S1}) that does not contain any JavaDoc (\eg the JavaDoc from lines 2-6 within \listref{lst:truncatenumber} is removed from the CUT). The second scenario (\textbf{S2}) has the JavaDoc but it does not include input/output examples,  only the method's behavior description (\eg \listref{lst:truncatenumber} will not have lines 4-5). The last scenario (\textbf{S3}) does not include the MUT's implementation, only its signature (\eg \listref{lst:truncatenumber} will not have lines 8-10).
%
\textbf{S1} and \textbf{S2} demonstrate the effect of changing  JavaDoc elements. Test-Driven Development (TDD)~\cite{beck2003TDD} inspires scenario \textbf{S3}, where test cases are written before the code implementation.

\noindent\textbf{\underline{SF110 Scenarios}}: Unlike HumanEval, the classes from SF110 do not necessarily include input/output pairs. Thus, we created scenarios slightly different than before. Scenario \textbf{S1} removes (i) any code within the class \textit{before} and \textit{after} the method under test as well as (ii) the class' JavaDoc. Scenario \textbf{S2} is the same as S1, but \textit{including} the JavaDoc for the method under test. Scenario \textbf{S3} is the same as S2, except that there is no method implementation for the MUT (only its signature). Scenario \textbf{S4} mimics S3, but it also includes all the fields and the signatures for the other methods/constructors in the MUT's declaring class.
Scenarios \textbf{S1} and \textbf{S2} demonstrate the effect of having or not having code documentation (JavaDoc).  \textbf{S3} verifies the usefulness of LLMs for TDD whereas \textbf{S4} is used to understand how code elements in a class are helpful for test generation.

After creating each of the scenarios above, we generated unit tests using the same models and following the steps outlined in \secref{sec:RQ1}. Then, we used JUnit5, JaCoCo, and \textsc{TsDetect} to measure test coverage, correctness, and quality. Similar to RQ1, we also compared the results to Evosuite~\cite{evosuite}.

 \section{RQ1 Results}\label{sec:resultRQ1}


We analyze the generated tests according to their:
\textbf{(i)} \textbf{\textit{compilation status}}; \textbf{(ii)} \textbf{\textit{correctness}}; \textbf{(iii)} \textbf{\textit{coverage}}; and \textbf{(iv)}  \textbf{\textit{quality}}.

\vspace{-5pt}
\subsection{Compilation Status}

\tblref{tab:rq1-compile} reports the percentage of generated unit tests that are compilable \textbf{\textit{before}} and \textbf{\textit{after}} applying the heuristic-based fixes described in \secref{subsec:DataAnalysis}.  
The number of unit tests and test methods for each model and dataset is shown in the last two columns of \tblref{tab:rq1-compile}. 
We obtained a total \textbf{2,536}  test methods (\ie a method with an \codeJava{@Test} annotation) scattered across \textbf{572} compilable Java test files for HumanEval and \textbf{2,022} test methods within \textbf{600} test files for SF110. 
For comparison, we also ran Evosuite~\cite{evosuite} (with default configuration parameters) to generate unit tests for each of the MUTs. Moreover, in the case of HumanEval, we manually created a JUnit5 test  for each input/output pair provided in each prompt.

\vspace{-5pt}
\subsubsection*{\underline{HumanEval Results}}\label{subsub:CompilationHumanEval}

On the one hand, we found that \finding{\textbf{\textit{less than half}} of the  unit tests generated by  \codexTwo, \codexFour, and \gpt{} are compilable for the classes in HumanEval}. 
On the other hand, \textbf{70\%} of \StarCoder's generated unit tests compiled. 
Upon applying heuristic-based fixes, the compilation rates have increased an average of \textbf{41\%}. The biggest increase was observed for the \codexTwo{} model; its compilation rate increased from \textbf{37.5\%} to \textbf{100\%}. \StarCoder{} was the LLM that the heuristics were the least able to improve; it only increased the compilation rate by \textbf{6.9\%}.


\vspace{-5pt}
\subsubsection*{\underline{SF110 Results}}\label{subsub:CompilationSF110}
For the SF110 dataset, the compilation rates are  lower than the ones observed for HumanEval. \finding{Between \textbf{2.7\%} and \textbf{12.7\%} of the generated unit tests for the SF110 dataset are compilable across all the studied LLMs}. 
\StarCoder{} was the LLM that generated the highest amount of compilable tests (\textbf{12.7\%}), whereas \codexTwo{} and \codexFour{} had the lowest compilation rate (\textbf{2.7\%} and \textbf{3.4\%}, respectively).
Similar to HumanEval, the heuristic-based fixes were able to increase the compilation rates by \textbf{81\%}, on average. Codex was the model with the highest increase; the compilation rates increased from less than \textbf{5\%} to over \textbf{99\%}. 
\StarCoder{} was the model that least benefited with our heuristics; its compilation rate increased by only \textbf{57.2\%}.

\vspace{-5pt}
\begin{table}[!ht]
\setlength{\tabcolsep}{2pt}
\centering
\caption{Compilation status of the generated unit tests}\label{tab:rq1-compile}\vspace{-10pt}
\scriptsize
\begin{tabular}{@{}cccccc@{}}
\toprule
& \textbf{LLM}            & \textbf{\begin{tabular}[c]{@{}c@{}}\% Compilable\end{tabular}} & \textbf{\begin{tabular}[c]{@{}c@{}}\% Compilable after fix\end{tabular}} & \textbf{\begin{tabular}[c]{@{}c@{}}\#Test Methods\end{tabular}}& \textbf{\begin{tabular}[c]{@{}c@{}}\#Test Classes\end{tabular}}\\ \midrule
\parbox[t]{3mm}{\multirow{6}{*}{\rotatebox[origin=c]{90}{\textbf{HumanEval}}}} 
& \textbf{GPT-3.5-Turbo}    & {43.1}\%  & {81.3}\%  & 1,117 & 130\\
& \textbf{StarCoder}     & {70.0}\%  & {76.9}\%  & 948 & 123 \\
& \textbf{Codex (2K)}     & {37.5}\%  & {100}\% & 697 & 160\\
& \textbf{Codex (4K)}     & {44.4}\%  & {99.4}\%  & 774 & 159\\ \cline{2-6}
& \textbf{Evosuite}       & 100\%  & NA  & 928 & 160\\
& \textbf{Manual}         & {100}\%  & NA  & 1,303 & 160\\\bottomrule
\parbox[t]{3mm}{\multirow{5}{*}{\rotatebox[origin=c]{90}{\textbf{SF110}}}} 
& \textbf{GPT-3.5-Turbo}    & {9.7}\%  & {85.9}\%  & 194 & 87\\
& \textbf{StarCoder}     & 12.7\%  & {69.8}\%  & 1,663 & 368 \\
& \textbf{Codex (2K)}     & {2.7}\%  & {74.5}\%  & 1,406 & 222\\
& \textbf{Codex (4K)}     & {3.4}\%  & {83.5}\%  & 1,039 & 152\\\cline{2-6}
& \textbf{Evosuite}       & 100\%  & NA  & 12,362 &	1,618\\\bottomrule
\end{tabular}
\end{table}
\vspace{-5pt}

\subsubsection*{\underline{Compilation error root causes}} 

The unit tests that were not fixable through heuristics were those that contained \textit{semantic} errors that failed the compilation. To observe the most common root causes of compilation errors, we collected all the compilation errors and clustered them using K-means~\cite{Lloyd1982LeastSQ}. We used the silhouette method~\cite{ROUSSEEUW198753} to find the number of clusters K ($K=48$). 
After inspecting these 48 clusters and making manual adjustments to clusters to fix imprecise clustering, we found that \finding{the top 3 compilation errors for \textsf{HumanEval} were caused by 
\textbf{\textit{unknown symbols}} (\ie the compiler cannot find the symbol), 
\textbf{\textit{incompatible conversion from \codeJava{java.util.List<T>} to \codeJava{java.util.List<X>}}}, and 
\textbf{\textit{incompatible conversion from \codeJava{int[]}  to \codeJava{java.util.List<Integer>}}}}. 
Unknown symbols accounted for more than \textbf{62\%} of the compilation errors. Several of these unknown symbols were caused by invoking non-existent methods or instantiating non-existent classes. For example, \StarCoder{} produced several test cases that invoked the method  \codeJava{java.util.List.of(int,int,int,...)}, which does not exist.
For the \textsf{SF110} dataset, \finding{the top 3 compilation errors were \textbf{\textit{unknown symbols}}, \textbf{\textit{class is abstract; cannot be instantiated}}, and \textbf{\textit{no suitable constructor found}}}.

\vspace{-5pt}
\subsection{Test Correctness}

We executed  each test that were compilable after our automated fix. We considered a unit test to be \textit{\textbf{correct}} if it had a success rate of \textbf{100\%} (\ie \textit{all} of its test methods passed) whereas a \textit{\textbf{somewhat correct}} unit test is one that had \textit{at least one} passing test method. As explained in ~\secref{subsec:DataAnalysis}, the reasoning behind these metrics is that the  HumanEval has a canonical solution which is the \textit{correct} implementation for the problem. Thus, a correct test must not fail (or else the input/output generated does not match the benchmark’s problem). Similarly, as the SF110 benchmark is a  popular benchmark for automatic test generation containing mature open-source projects, they have a higher probability that they are functionally correct. Both metrics are reported in \tblref{tab:rq1-correct}.

\vspace{-5pt}
\subsubsection*{\underline{HumanEval Results}}
\finding{\StarCoder{} generated the highest amount of correct unit tests ($\approx$81\%)}. Although \gpt{} only produced \textbf{52\%} correct unit tests, it was the model that generated the highest amount of tests that have at \textit{at least one} passing test method (\textbf{92.3\%}).  We also found that increasing Codex's token size did not yield higher correctness rates.
Moreover, between \textbf{52\%} to \textbf{81\%} of generated tests were correct whereas  \textbf{81\%}-\textbf{92\%} of the tests had \textit{at least one} passing test case.
From these results, we can infer that although all the models could not produce correct tests, they can still be useful in generating at least a few viable input/output pairs.

\begin{table}[ht]
\setlength{\tabcolsep}{5pt}
\centering
\caption{Correct tests percentage for HumanEval and SF110}
\label{tab:rq1-correct}\vspace{-10pt}
\scriptsize\centering
\begin{tabular}{@{}crcccc@{}}
\toprule
    & & \textbf{GPT-3.5-Turbo} & \textbf{StarCoder} & \textbf{\begin{tabular}[c]{@{}c@{}}Codex (2K)\end{tabular}} & \textbf{\begin{tabular}[c]{@{}c@{}}Codex (4K)\end{tabular}} \\ \midrule

\parbox[t]{2mm}{\multirow{2}{*}{\rotatebox[origin=c]{90}{\tiny\textbf{HE}}}}
    & \textbf{\begin{tabular}[c]{@{}r@{}}\% Correct\end{tabular}}   
    & 52.3\% & 81.3\% & 77.5\%& 76.7\% \\
    & \textbf{\begin{tabular}[c]{@{}r@{}}\% Somewhat Correct\end{tabular}} 
    & 92.3\% & 81.3\% & 87.5\%& 87.4\%         \\ \bottomrule

\parbox[t]{2mm}{\multirow{2}{*}{\rotatebox[origin=c]{90}{\tiny\textbf{SF110}}}}
    & \textbf{\begin{tabular}[c]{@{}r@{}}\% Correct\end{tabular}}   
    & 6.9\% & 51.9\% & 46.5\% & 41.1\% \\
    & \textbf{\begin{tabular}[c]{@{}r@{}}\% Somewhat Correct\end{tabular}} 
    & 16.1\% & 58.6\% & 62.7\% & 53.7\%         \\ \bottomrule

\end{tabular}
\end{table}

\vspace{-5pt}
\subsubsection*{\underline{SF110 Results}}
The  correctness rates achieved by the LLMs are rather low. \finding{Less than \textbf{52\%} of the produced tests are  correct for all models}.
Even when considering the unit tests that produced at least one passing test case (\textit{somewhat correct}), only up to \textbf{63\%}  fulfill this criterion.  The best-performing model for the SF110 dataset was \StarCoder{}, which produced \textbf{51.9\%} correct tests.   \codexTwo{} was the best performing LLM for generating unit tests that have \textit{at least one}  passing test case.

\subsection{Test Coverage}

\vspace{-1pt}
\subsubsection*{\underline{HumanEval Results}}
\tblref{tab:rq1-coverage} shows the line and branch coverage for the HumanEval dataset, computed considering all the Java classes in the dataset. The LLMs achieved line coverage ranging from \textbf{67\%} to \textbf{87.7\%} and branch coverage ranging from \textbf{69.3\%} to \textbf{92.8\%}. \finding{Codex (4K) exhibited the highest line and branch coverage of \textbf{87.7\%} and \textbf{92.8\%}, respectively}. However, \finding{the coverage of the unit tests generated by LLMs are below the coverage reported by the manual tests and those generated by Evosuite}. In fact, Evosuite, which relies on an evolutionary algorithm to generate JUnit tests, has a higher line and branch coverage than the manually written tests. 

\begin{table}[!ht]
\centering
\setlength{\tabcolsep}{3pt}
\caption{Line and branch coverage}
\label{tab:rq1-coverage}\vspace{-10pt}
\scriptsize\centering
\begin{tabular}{@{}crcccccc@{}}
\toprule
& \multicolumn{1}{c}{\textbf{Metric}}                                & \textbf{GPT-3.5-Turbo} & \textbf{StarCoder} & \textbf{\begin{tabular}[c]{@{}c@{}}Codex-2K\end{tabular}} & \textbf{\begin{tabular}[c]{@{}c@{}}Codex-4K\end{tabular}} & \textbf{Evosuite} & \textbf{Manual} \\ \midrule
\parbox[t]{1.8mm}{\vspace{-6pt}\multirow{3}{*}{\rotatebox[origin=c]{90}{\tiny\textbf{HumanEval}}}}
    &
\textbf{\begin{tabular}[c]{@{}r@{}}Line\\ Coverage\end{tabular}}   
& 69.1\%               & 67.0\%           & 87.4\%                                                        & 87.7\%                                                        & 96.1\%            & 88.5\%          \\
& \textbf{\begin{tabular}[c]{@{}r@{}}Branch\\ Coverage\end{tabular}} & 76.5\%               & 69.3\%           & 92.1\%                                                        & 92.8\%                                                        & 94.3\%            & 93.0\%          \\ \hline
\parbox[t]{1.8mm}{\multirow{3}{*}{\rotatebox[origin=c]{90}{\tiny\textbf{SF110}}}}
    & \textbf{\begin{tabular}[c]{@{}r@{}}Line\\ Coverage\end{tabular}}   
    & 1.3\% & 1.1\% & 1.9\% & 1.2\% & 27.5\% & --\\
    & \textbf{\begin{tabular}[c]{@{}r@{}}Branch\\ Coverage\end{tabular}}    
    & 1.6\% & 0.5\% & 1.1\% & 0.7\% & 20.2\% & --
\\ \bottomrule
\end{tabular}
\end{table}

\vspace{-5pt}
\subsubsection*{\underline{SF110 Results}}
The test coverage for SF110 is worse when compared to HumanEval (they were less than \textbf{2\%}  for all models). \codexTwo{} was the best performing one in terms of line coverage (\textbf{1.9\%}), whereas \gpt{} had the highest  branch coverage (\textbf{1.6\%}). Yet, these coverages are  $\approx$11-19$\times$ lower than the coverage achieved by Evosuite's tests. 

\subsection{Test Smells}

\vspace{-5pt}
\subsubsection*{\underline{HumanEval Results}}
\tblref{tab:rq1-humaneval-smells} shows that the LLMs produced the following smells\footnote{We hide 
\textit{Default Test}, \textit{General Fixture}, \textit{Mystery Guest}, \textit{Verbose Test}, \textit{Resource Optimism},  \textit{Dependent Test}, and other test smell types supported by \textsc{TsDetect} because they did not occur in any of the listed approaches}: 
\textsf{Assertion Roulette (AR)}~\cite{Simula.SE.525},
\textsf{Conditional Logic Test (CLT)}~\cite{meszaros03},
\textsf{Empty Test (EM)}~\cite{peruma19}, 
\textsf{Exception Handling (EH)}~\cite{peruma19},
\textsf{Eager Test (EA)}~\cite{Simula.SE.525},
\textsf{Lazy Test (LT)}~\cite{Simula.SE.525},
\textsf{Duplicate Assert (DA)}~\cite{peruma19},
\textsf{Unknown Test (UT)}~\cite{peruma19},
, and \textsf{Magic Number Test (MNT)}~\cite{meszaros03}.
We found that \finding{Magic Number Test (MNT) and Lazy Test (LT) are the two most reoccurring test smell types across \emph{all} the approaches, \ie in the unit tests generated by the LLMs and Evosuite as well as the ones created manually}. The \textbf{MNT} smell occurs when the unit test hard-codes a value in an assertion without a comment explaining it, whereas the \textbf{LT} smell arises when multiple test methods invoke the same production code. 


\begin{table}[!ht]
\centering
\newcommand{\midsepremove}{\aboverulesep = 0mm \belowrulesep = 0mm}
\midsepremove
\setlength{\tabcolsep}{3.1pt}
\caption{Test smells distribution for the HumanEval dataset.}
\label{tab:rq1-humaneval-smells}\vspace{-10pt}
\scriptsize
\begin{tabular}{@{}rcccccc@{}}
\toprule
\textbf{Test Smell} & \textbf{\begin{tabular}[c]{@{}c@{}}Codex (2K)\end{tabular}} & \textbf{\begin{tabular}[c]{@{}c@{}}Codex (4K)\end{tabular}} & \textbf{StarCoder} & \textbf{GPT-3.5-Turbo} & \textbf{Evosuite} & \textbf{Manual} \\ \midrule
\textbf{AR}  & 61.3\%                                                        & 59.7\%                                                        & 51.3\%           & 23.8\%               & 15.0\%            & 0.0\%           \\
\textbf{CLT} & 0.0\%                                                         & 0.0\%                                                         & 0.0\%            & 1.5\%                & 0.0\%             & 0.0\%           \\
\textbf{EM}  & 1.9\%                                                         & 1.3\%                                                         & 3.8\%            & 0.8\%                & 0.0\%             & 0.0\%           \\
\textbf{EH}  & 0.0\%                                                         & 0.0\%                                                         & 0.0\%            & 0.0\%                & 100.0\%           & 100.0\%         \\
\textbf{EA}  & 60.6\%                                                        & 59.1\%                                                        & 48.8\%           & 23.8\%               & 16.3\%            & 0.0\%           \\
\textbf{LT}  & 39.4\%                                                        & 41.5\%                                                        & 51.3\%           & 86.2\%               & 99.4\%            & 100.0\%         \\
\textbf{DA}  & 15.6\%                                                        & 14.5\%                                                        & 10.6\%           & 3.1\%                & 0.6\%             & 0.0\%           \\
\textbf{UT}  & 10.0\%                                                        & 5.7\%                                                         & 6.3\%           & 0.8\%                & 0.0\%             & 0.0\%           \\
\textbf{MNT} & 100.0\%                                                       & 100.0\%                                                       & 100\%           & 100.0\%              & 100.0\%           & 100.0\%        \\ \bottomrule
\end{tabular}
\end{table}

\finding{Whereas Codex, StarCoder, and GPT-3.5-Turbo did not produce unit tests with the \textsf{Exception Handling} (EH) smell, this smell type was frequent in all manually created tests and those generated by Evosuite}. 
We also found that \finding{\textsf{Assertion Roulette (AR)}  is a common smell produced by LLMs (frequency between \textbf{23.8\%} -- \textbf{61.3\%}) and that also occurred in Evosuite in \textbf{15\%} of its generated tests}. This smell occurs when the same test method invokes an \code{assert} statement to check for different input/output pairs and does not include an error message for each of these asserts. Similarly, \finding{the LLMs and Evosuite also produced unit tests with the \textsf{Eager Test smell (EA)}}, in which a single test method invokes  different methods from the production class, as well as the \textsf{Duplicate Assert smell (DA)} (caused by multiple assertions for the same input/output pair).


\vspace{-5pt}
\subsubsection*{\underline{SF110 Results}}
The smells detected for the SF110 tests are listed in \tblref{tab:rq1-sf110-smells}. Similar to HumanEval, \textsf{Magic Number Test (MNT)}, \textsf{Assertion Roulette (AR)}, and \textsf{Eager Tests (EA)} are frequently occurring smells in the unit tests generated by the LLMs and Evosuite. 
The LLMs generated other types of smells  that were not observed for the HumanEval dataset, namely 
\textsf{Constructor Initialization (CI)}~\cite{peruma19},
\textsf{Mistery Guest (MG)}~\cite{Simula.SE.525},
\textsf{Redundant Print (RP)}~\cite{peruma19},
\textsf{Redundant Assertion (RA)}~\cite{peruma19},
\textsf{Sensitive Equality (SE)}~\cite{Simula.SE.525},
\textsf{Ignored Test (IT)}~\cite{peruma19}, and
\textsf{Resource Optimism (RO)}~\cite{peruma19}.

While LLMs produced tests that had  \textsf{Empty Tests (EM)}, \textsf{Redundant Print (RP)}, \textsf{Redundant Assertion (RA)}, and \textsf{Constructor Initialization (CI)}  smells,  Evosuite did not generate any unit test with these smell types. 
We also observed that \StarCoder{} generated (proportionally) more samples than the other models (\textbf{96.7\%} of its generated tests had at least one test smell).

\begin{table}[!ht]
\centering
\newcommand{\midsepremove}{\aboverulesep = 0mm \belowrulesep = 0mm}
\midsepremove
\setlength{\tabcolsep}{6.5pt}
\caption{Test smells distribution for the SF110 dataset (RQ1).}
\label{tab:rq1-sf110-smells}\vspace{-10pt}
\scriptsize
\begin{tabular}{@{}cccccc@{}}
\toprule
\textbf{Test Smell} & \textbf{\gpt} & \textbf{\StarCoder} & \textbf{Codex (2K)} & \textbf{Codex (4K)} &   \textbf{Evosuite}  \\ \midrule
\textbf{AR} &  4.6\% &  35.1\% &  14.4\% &  17.1\% &  35.0\% \\
\textbf{CLT} &  2.3\% &  2.4\% &  0.5\% &  1.3\% &  0.0\% \\
\textbf{CI} &  0.0\% &  4.9\% &  0.0\% &  0.7\% &  0.1\% \\
\textbf{EM} &  0.0\% &  3.8\% &  7.2\% &  1.3\% &  0.0\% \\
\textbf{EH} &  2.3\% &  18.2\% &  20.7\% &  19.1\% &  91.2\% \\
\textbf{MG} &  0.0\% &  3.5\% &  2.7\% &  3.3\% &  3.0\% \\
\textbf{RP} &  0.0\% &  10.6\% &  4.5\% &  5.9\% &  0.0\% \\
\textbf{RA} &  0.0\% &  0.3\% &  0.9\% &  0.7\% &  0.0\% \\
\textbf{SE} &  0.0\% &  1.9\% &  0.9\% &  1.3\% &  13.7\% \\
\textbf{EA} &  12.6\% &  39.7\% &  28.4\% &  31.6\% &  39.6\% \\
\textbf{LT} &  21.8\% &  33.4\% &  60.8\% &  60.5\% &  46.4\% \\
\textbf{DA} &  1.1\% &  11.7\% &  1.4\% &  2.0\% &  1.5\% \\
\textbf{UT} &  0.0\% &  21.2\% &  21.2\% &  10.5\% &  22.9\% \\
\textbf{IT} &  0.0\% &  0.3\% &  0.0\% &  0.0\% &  0.0\% \\
\textbf{RO} &  0.0\% &  4.6\% &  2.7\% &  3.9\% &  2.7\% \\
\textbf{MNT} &  21.8\% &  95.4\% &  93.2\% &  96.1\% &  91.2\% \\\bottomrule
\end{tabular}
\end{table}

 \section{RQ2 Results}\label{sec:resultRQ2}

Similar to RQ1, we investigated how code elements in a context influence the generated unit tests with respect to their \textit{\textbf{compilation status}}, \textit{\textbf{correctness}}, \textit{\textbf{coverage}}, and \textit{\textbf{quality}}.

\vspace{-5pt}
\subsection{Compilation Status}

\begin{figure}[!h]
    \centering
  \subfloat{\includegraphics[width=.44\linewidth]{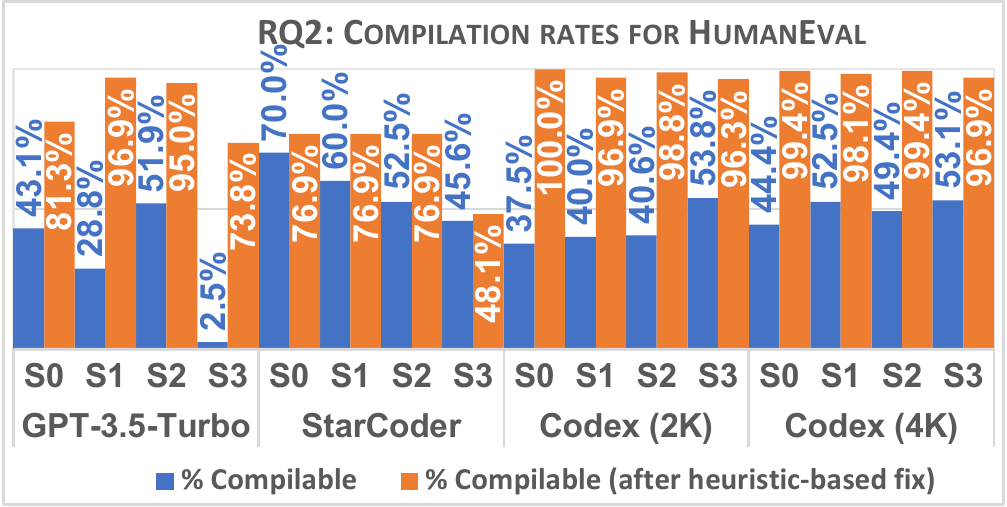}}
  \subfloat{\includegraphics[width=.56\linewidth]{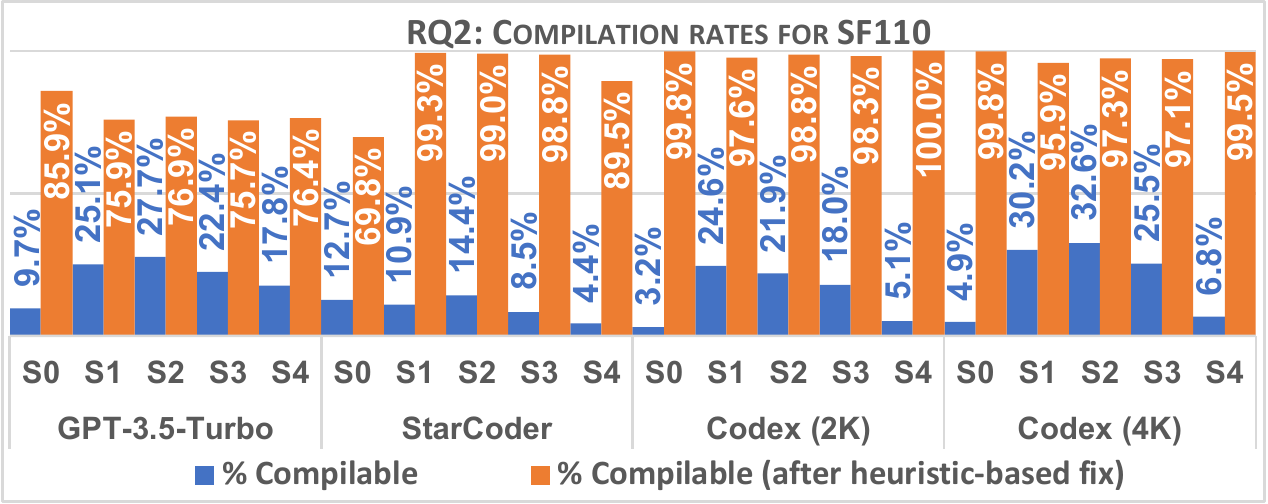}}
  \vspace{-10pt}\caption{Compilation rates for HumanEval and SF110}\label{fig:rq2-compilation}
\end{figure}

\figref{fig:rq2-compilation} shows the compilation rates for the HumanEval and SF110 datasets across the different scenarios and LLMs. 

\vspace{-10pt}
\vspace{-10pt}

\subsubsection*{\underline{HumanEval Results}}
 \finding{Scenario 3 (\textbf{\textsf{S3}}) increased the original (\textbf{\textsf{S0}}) compilation rates for Codex (2K and 4K) from \textbf{37.5\%}, and \textbf{44.4\%} to \textbf{53.8\%} and \textbf{53.1\%}, respectively. Although scenario 3 increased the original compilation rates (blue bars in \figref{fig:rq2-compilation}), these tests have similar heuristic-based fix rates. 
In the case of \StarCoder, the original prompt (\textbf{\textsf{S0}}) was the best in generating compilable code.
GPT-3.5-Turbo, on the other hand, experienced a sharp decrease from 43.1\% to 2.5\%} for S3. Upon further inspection, we found that scenario 3 triggered GPT-3.5-Turbo 3.5 to include the original class under test in its entirety, followed by the unit test. This resulted in two package declarations on the produced output; one was placed in the very first line (corresponding to the CUT's package), and the other was placed after the CUT for the unit test's package. These duplicated package declarations lead to compilation errors. These issues were later fixed by applying the heuristic \textbf{H3}. For the \gpt{} model, the best-performing context scenario was \textbf{\textsf{S2}}, in which the prompt does not include sample input/output pairs.

\vspace{-5pt}
\subsubsection*{\underline{SF110 Results}}

\textbf{\textsf{S2}} increased the original (\textbf{\textsf{S0}}) compilation rates for GPT-3.5-Turbo, StarCoder, and Codex (4K). However, scenario 1 (\textbf{\textsf{S1}}) was the best performer for Codex (2K), while scenario 2 (\textbf{\textsf{S2}}) was the second-best performer.
What these results show is that it is beneficial to include a \textit{minimal} context, which contains only  the MUT's implementation when generating test cases. The benefit seems twofold: (1) it can increase the compilation rate of the generated code snippets, and (2) it consumes less input tokens, as other methods from the class under test are removed.

\vspace{-5pt}
\subsection{Test Correctness}
\begin{figure}[!h]
  \subfloat{\includegraphics[width=.445\linewidth]{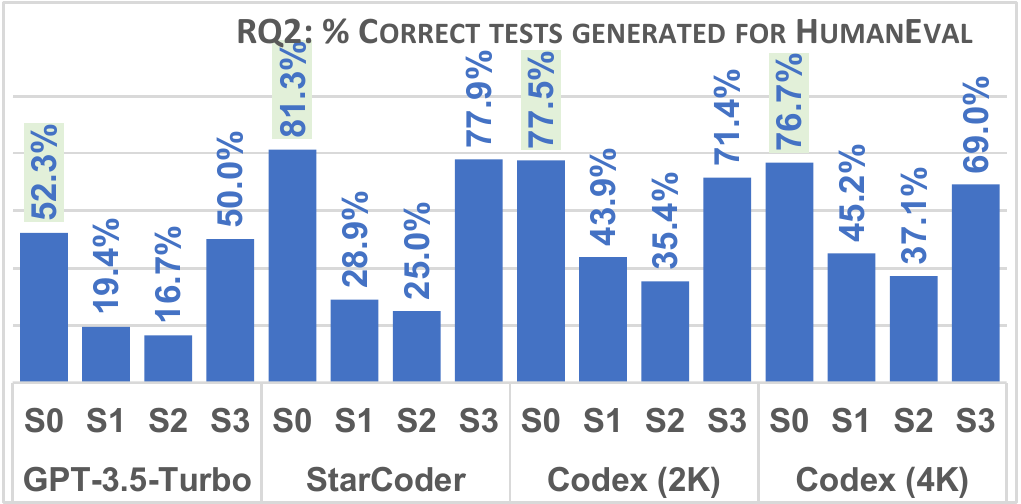}}
  \subfloat{\includegraphics[width=.555\linewidth]{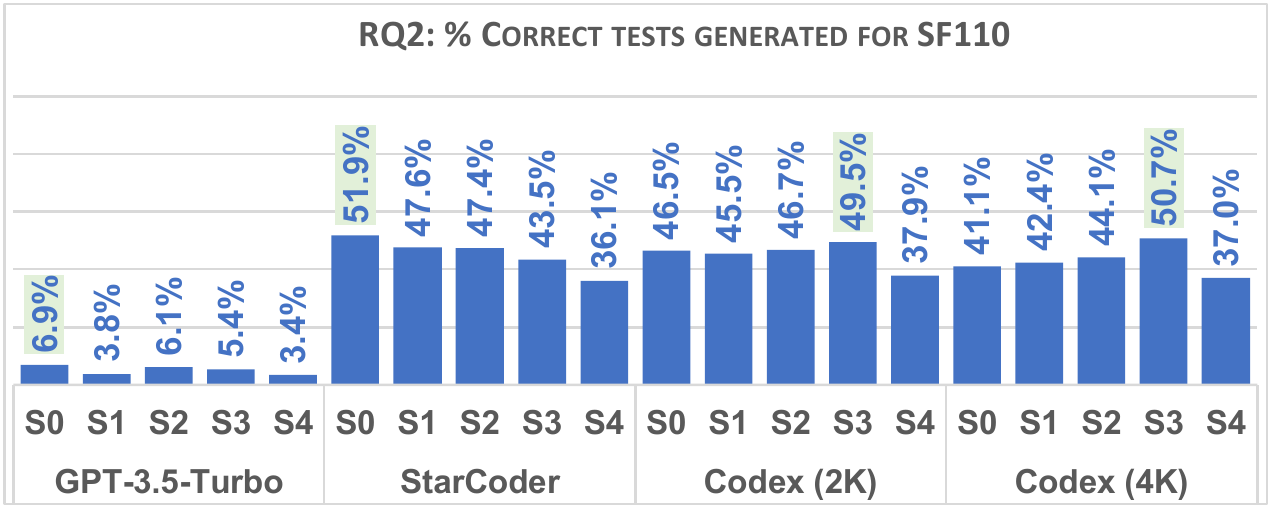}}
  \vspace{-10pt}\caption{Correctness rates}\label{fig:rq2-correctness}
\end{figure}

\figref{fig:rq2-correctness} shows the percentage of unit tests generated by the LLMs that are \textit{correct} for the HumanEval and SF110 datasets. The best performing scenarios for an LLM are highlighted in green.

\vspace{-10pt}
\vspace{-10pt}

\subsubsection*{\underline{HumanEval Results}}
The original context (\textbf{\textsf{S0}}) is the one that leads to the highest amount of correct tests for the HumanEval dataset.  Among all scenarios, \finding{scenario 3 \textbf{\textsf{S3}})  had a \textit{similar} correctness rate compared to the original prompt used in RQ1 for GPT-3.5-Turbo and Codex (2K, 4K)}. 
It is important to highlight that  whereas GPT-3.5-Turbo only had 73.8\% compilable tests in scenario 3 (compared to 81.3\% tests from the original prompt) it still had a similar correctness rate. Yet, the original prompt is the one that has the highest correctness rates. 
Recall that scenario 3 (\textbf{\textsf{S3}}) is the one in which the implementation of the method under test is not included in the prompt. These results show that LLMs can still generate unit tests even if the implementation is not provided. Such a scenario can be useful in TDD; where developers write tests \textit{before} the production code.

\paragraph{\bf $\bullet$ Effects on including input/output examples on the prompt}

The HumanEval dataset has input/output examples in its problem description (see ~\listref{lst:truncatenumber}). Thus, for this dataset, we also investigated to what extent LLMs are able to generate unique input/output pairs that are not included in the original problem description and how these are related to the test correctness rates observed. 
We manually inspected each generated test to compute the \textit{\textbf{total number of unique input/output pairs}} generated.
For each unique input/output pair, we compared with the ones provided in the problem's description in order to compute the \textit{\textbf{total number of input-output pairs that are from the problem description}} and the \textit{\textbf{total number of input-output pairs that are \emph{\underline{not}} in the problem description}}.

\begin{figure}[!h]
    \centering
  \subfloat{\includegraphics[width=.53\linewidth]{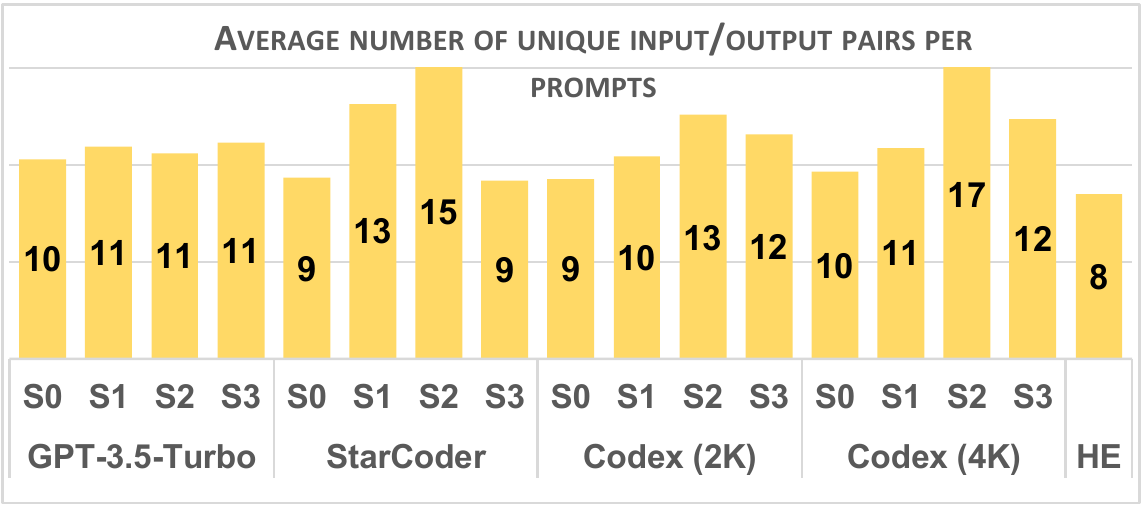}}
  \subfloat{\includegraphics[width=.47\linewidth]{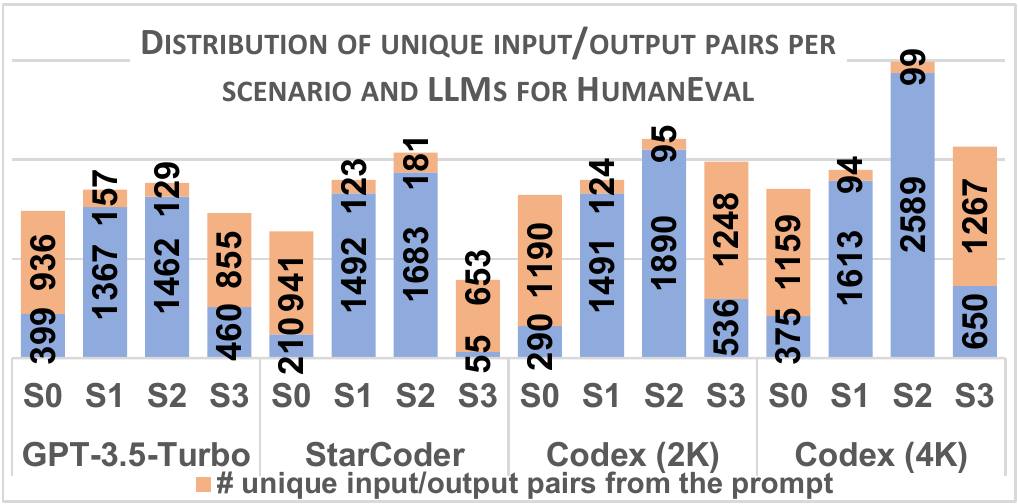}}\vspace{-10pt}
  \caption{\textit{Left}: Average number of unique input/outputs per prompt for each LLM and the original dataset (HumanEval - HE). \textit{Right}: Total number of unique input/outputs that are and are not from the problem's description.}\label{fig:HumanEvalParroting}
\end{figure}

\figref{fig:HumanEvalParroting} (\textit{left}) shows the average number of unique input/output pairs for each LLM and scenario combination compared to the problem description in the HumanEval dataset. Each problem in the HumanEval dataset  provides an average of \textbf{8} input/output pair examples, whereas the LLMs provide more than that, as the prompts explicitly request 10 test cases for each problem description. We observed that the scenarios \textbf{\textsf{S1}} and \textbf{\textsf{S2}}, which do not include input-output pairs in the prompt, has a \textit{higher} average of a number of unique input-output pairs.

\figref{fig:HumanEvalParroting} (right)  shows how many of the generated input/output pairs by the LLMs are from the problem's description and how many are not.  We found that the scenarios  \textbf{\textsf{S1}} and \textbf{\textsf{S2}} generated \textit{more} input-output pairs that are not from the original description, whereas the scenarios \textbf{\textsf{S0}} and \textbf{\textsf{S3}} are repeating the test cases from the prompt.
That is, the models are behaving like ``parrots''~\cite{bender2021dangers} by using the same input/output in the prompt and just formatting it as a test case without generating new examples. 
When contrasting with the correctness rates observed in ~\figref{fig:rq2-correctness} we see that scenarios \textbf{\textsf{S1}} and \textbf{\textsf{S2}} were consistently lower for all LLMs. These results show that although scenarios \textbf{\textsf{S1}} and \textbf{\textsf{S2}}  generated \textit{more} input-output examples, those were not necessarily correct. The prompts that included examples of input-outputs had \textit{higher} correctness rates.


\vspace{-5pt}
\subsubsection*{\underline{SF110 Results}}

While the original prompt (\textbf{\textsf{S0}})  achieved the highest correctness rate for GPT-3.5-Turbo (6.9\%) and StarCoder (51.9\%), the other LLMs observed a correctness increase when using the context from scenario 3 (\textbf{\textsf{S3}}).  Codex (4K) experienced the highest increase (from 37.9\% to 50.7\%) for S3.
This scenario (\textbf{\textsf{S3}}) has a context that only includes the MUT's Javadoc and signature and removes other methods from the class where the MUT is declared.

\vspace{-10pt}
\subsection{Test Coverage}

\begin{figure}[H]\centering
  \subfloat{\includegraphics[width=.45\linewidth]{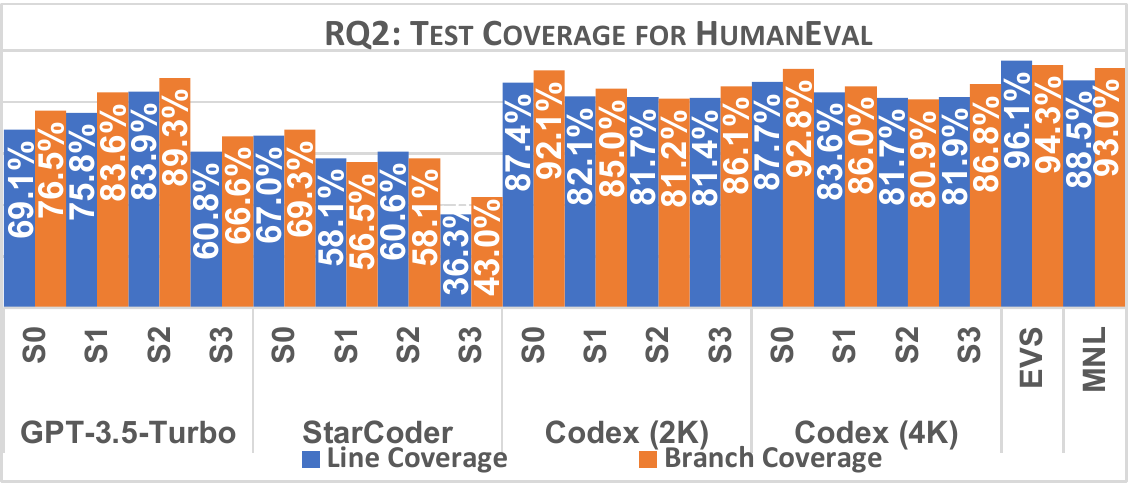}}
  \subfloat{\includegraphics[width=.54\linewidth]{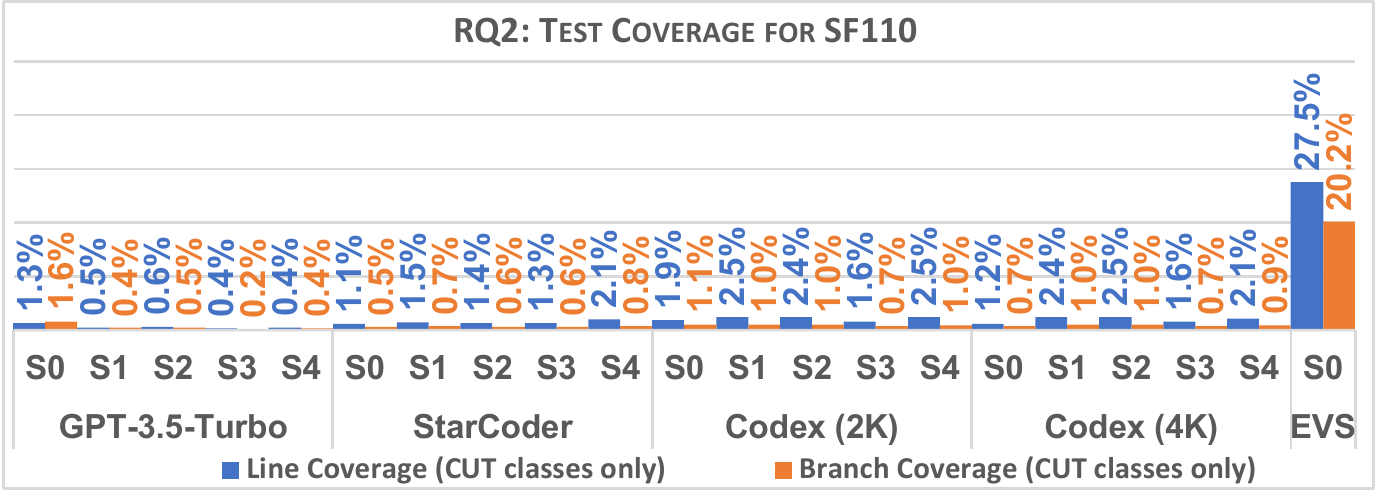}}\vspace{-10pt}
  \caption{Line and Branch Coverage across different datasets, scenarios, and LLMs (EVS = Evosuite; MNL = Manual).}\label{fig:rq2-coverage}
\end{figure}

\figref{fig:rq2-coverage} shows the \textit{line} and \textit{branch} coverage for each scenario.

\vspace{-10pt}
\vspace{-10pt}

\subsubsection*{\underline{HumanEval Results}}

For Codex, scenario 1 is the one that had the highest line coverage among the different scenarios in these models. GPT-3.5-Turbo and StarCoder, on the other hand, had scenario 2 as the one with the highest \textit{line} coverage. 
With respect to \textit{branch} coverage, we found that scenario 3 was the best performing one for Codex, and scenario 2 is the best one for GPT-3.5-Turbo and StarCoder. None of the scenarios for Codex (2K and 4K) and StarCoder outperformed the line/branch coverage of the original prompts nor the coverage achieved by the manual and Evosuite's tests.

\subsubsection*{\underline{SF110 Results}}

Among all scenarios, \finding{scenario 1 (\textbf{\textsf{S1}}) and scenario 2 (\textbf{\textsf{S2}}) had a slightly \textit{higher} line coverage when compared to the original prompt (\textbf{S0}) used in RQ1 for Codex (2K) and Codex (4K), respectively}. For StarCoder the scenario 4 had a higher line coverage than the original one. The original context of \gpt{}, on the other hand, had the highest observed line coverage.
In the case of branch coverage, scenario 1 (\textbf{\textsf{S1}}) had slightly higher coverage for \codexFour, whereas scenario 4 (\textbf{\textsf{S4}}) was the best one for StarCoder. However, these increases  are still much lower than Evosuite's test coverage, which achieved $\approx$ 27\% line and branch coverage.

\vspace{-5pt}
\subsection{Test Smells}

\subsubsection*{\underline{HumanEval Results}}
\tblref{tab:rq2-humaneval-smells} shows the distribution of smells for different scenarios and LLMs. The cells highlighted in \colorbox[HTML]{DDFFDD}{green} are those in which the percentage is lower than the original context, whereas those highlighted in \colorbox[HTML]{FFCCC9}{red} have a higher percentage than the original context. In terms of smell types, \finding{all scenarios have the same smell types that occurred  in the original prompts (see \tblref{tab:rq1-humaneval-smells})}. 
We also observe that, overall, \finding{the scenarios tended to decrease the incidence of generated smells}.
When comparing each scenario to one another, \finding{there is no clear outperformer across all the LLMs}. Yet, Scenario 3 for GPT-3.5-Turbo had higher percentages than the original context, on average. Although the average increases are not significant (0.6\% and 0.2\% for these LLMs, respectively).

\vspace{-10pt}
\begin{table}[!ht]
\centering
\newcommand{\midsepremove}{\aboverulesep = 0mm \belowrulesep = 0mm}
\midsepremove
\setlength{\tabcolsep}{2.5pt}
\caption{Test smells distribution for the HumanEval dataset.}\label{tab:rq2-humaneval-smells}\vspace{-10pt}
\scriptsize
\begin{tabular}{@{}ccccccccccccc@{}}
\toprule
 & \multicolumn{3}{c}{\textbf{GPT-3.5-Turbo}} & 
 \multicolumn{3}{c}{\textbf{StarCoder}} & 
 \multicolumn{3}{c}{\textbf{Codex (2K)}} & 
 \multicolumn{3}{c}{\textbf{Codex (4K)}} \\ 
 & \textbf{S1} & \textbf{S2} & \textbf{S3} & 
 \textbf{S1} & \textbf{S2} & \textbf{S3} & 
 \textbf{S1} & \textbf{S2} & \textbf{S3} & 
 \textbf{S1} & \textbf{S2} & \textbf{S3} \\
 \midrule
\textbf{AR} & \cellcolor[HTML]{F3FFF3}7.1\% & \cellcolor[HTML]{F3FFF3}11.8\% & \cellcolor[HTML]{FFF4F3}30.5\% & \cellcolor[HTML]{F3FFF3}36.9\% & \cellcolor[HTML]{F3FFF3}36.3\% & \cellcolor[HTML]{F3FFF3}48.1\% & \cellcolor[HTML]{F3FFF3}16.8\% & \cellcolor[HTML]{F3FFF3}38.6\% & \cellcolor[HTML]{F3FFF3}61.0\% & \cellcolor[HTML]{F3FFF3}16.6\% & \cellcolor[HTML]{F3FFF3}40.3\% & \cellcolor[HTML]{FFF4F3}63.2\% \\
\textbf{CLT} & \cellcolor[HTML]{FFF4F3}6.5\% & \cellcolor[HTML]{FFF4F3}3.3\% & \cellcolor[HTML]{F3FFF3}0.8\% & 0.0\% & \cellcolor[HTML]{FFF4F3}0.6\% & 0.0\% & 0.0\% & 0.0\% & 0.0\% & 0.0\% & 0.0\% & 0.0\% \\
\textbf{EM} & \cellcolor[HTML]{F3FFF3}0.0\% & \cellcolor[HTML]{F3FFF3}0.7\% & \cellcolor[HTML]{FFF4F3}3.4\% & \cellcolor[HTML]{F3FFF3}1.9\% & \cellcolor[HTML]{FFF4F3}8.1\% & 3.8\% & \cellcolor[HTML]{FFF4F3}4.5\% & \cellcolor[HTML]{FFF4F3}3.2\% & \cellcolor[HTML]{FFF4F3}1.9\% & \cellcolor[HTML]{F3FFF3}1.3\% & \cellcolor[HTML]{F3FFF3}1.3\% & \cellcolor[HTML]{F3FFF3}1.9\% \\
\textbf{EA} & \cellcolor[HTML]{F3FFF3}7.1\% & \cellcolor[HTML]{F3FFF3}10.5\% & \cellcolor[HTML]{FFF4F3}26.3\% & \cellcolor[HTML]{F3FFF3}28.8\% & \cellcolor[HTML]{F3FFF3}30.0\% & \cellcolor[HTML]{F3FFF3}48.1\% & \cellcolor[HTML]{F3FFF3}15.5\% & \cellcolor[HTML]{F3FFF3}37.3\% & \cellcolor[HTML]{F3FFF3}56.5\% & \cellcolor[HTML]{F3FFF3}15.3\% & \cellcolor[HTML]{F3FFF3}38.4\% & \cellcolor[HTML]{FFF4F3}58.1\% \\
\textbf{LT} & \cellcolor[HTML]{F3FFF3}85.2\% & \cellcolor[HTML]{FFF4F3}92.8\% & \cellcolor[HTML]{F3FFF3}82.2\% & \cellcolor[HTML]{FFF4F3}61.9\% & \cellcolor[HTML]{FFF4F3}63.8\% & \cellcolor[HTML]{FFF4F3}53.1\% & \cellcolor[HTML]{FFF4F3}84.5\% & \cellcolor[HTML]{FFF4F3}60.8\% & \cellcolor[HTML]{FFF4F3}44.2\% & \cellcolor[HTML]{FFF4F3}84.7\% & \cellcolor[HTML]{FFF4F3}60.4\% & \cellcolor[HTML]{F3FFF3}42.6\% \\
\textbf{DA} & \cellcolor[HTML]{F3FFF3}1.3\% & \cellcolor[HTML]{F3FFF3}0.0\% & \cellcolor[HTML]{F3FFF3}1.7\% & \cellcolor[HTML]{F3FFF3}8.1\% & \cellcolor[HTML]{FFF4F3}11.3\% & \cellcolor[HTML]{FFF4F3}11.3\% & \cellcolor[HTML]{F3FFF3}0.6\% & \cellcolor[HTML]{F3FFF3}8.2\% & \cellcolor[HTML]{F3FFF3}11.0\% & \cellcolor[HTML]{F3FFF3}1.9\% & \cellcolor[HTML]{F3FFF3}6.9\% & \cellcolor[HTML]{FFF4F3}11.6\% \\
\textbf{UT} & \cellcolor[HTML]{F3FFF3}0.0\% & \cellcolor[HTML]{F3FFF3}0.7\% & \cellcolor[HTML]{FFF4F3}3.4\% & \cellcolor[HTML]{FFF4F3}11.3\% & \cellcolor[HTML]{FFF4F3}13.8\% & 6.3\% & \cellcolor[HTML]{FFF4F3}13.5\% & \cellcolor[HTML]{FFF4F3}16.5\% & \cellcolor[HTML]{F3FFF3}2.6\% & \cellcolor[HTML]{F3FFF3}5.1\% & \cellcolor[HTML]{FFF4F3}8.2\% & \cellcolor[HTML]{F3FFF3}2.6\% \\
\textbf{MNT} & \cellcolor[HTML]{F3FFF3}89.7\% & \cellcolor[HTML]{F3FFF3}98.7\% & 100\% & \cellcolor[HTML]{F3FFF3}99.4\% & \cellcolor[HTML]{F3FFF3}99.4\% & 100\% & 100\% & 100\% & 100\% & 100\% & 100\% & 100\% \\ \bottomrule
\end{tabular}
\end{table}

\vspace{-5pt}
\subsubsection*{\underline{SF110 Results}}

As shown \tblref{tab:rq2-sf110-smells}, there is not any scenario that consistently outperforms the other. However, we can observe that scenario 2 for GPT-3.5-Turbo produces more test smells than the other scenarios, as we can see from the 
cells highlighted in \colorbox[HTML]{FFCCC9}{red}. 

\vspace{-10pt}
\begin{table}[!ht]
\centering
\newcommand{\midsepremove}{\aboverulesep = 0mm \belowrulesep = 0mm}
\midsepremove
\setlength{\tabcolsep}{0.45pt}
\caption{Test smells distribution for the SF110 dataset (RQ2).}\label{tab:rq2-sf110-smells}\vspace{-10pt}
\scriptsize
\begin{tabular}{@{}ccccccccccccccccc@{}}
\toprule
 & \multicolumn{4}{c}{\textbf{Codex (2K)}} & \multicolumn{4}{c}{\textbf{Codex (4K)}} & \multicolumn{4}{c}{\textbf{StarCoder}} & \multicolumn{4}{c}{\textbf{GPT-3.5-Turbo}} \\ 
 & \textbf{S1} & \textbf{S2} & \textbf{S3} & \textbf{S4} & \textbf{S1} & \textbf{S2} & \textbf{S3} & \textbf{S4} & \textbf{S1} & \textbf{S2} & \textbf{S3} & \textbf{S4} & \textbf{S1} & \textbf{S2} & \textbf{S3} & \textbf{S4} \\\midrule
\textbf{AR} & \cellcolor[HTML]{FFF4F3}17.3\% & \cellcolor[HTML]{F3FFF3}12.8\% & \cellcolor[HTML]{F3FFF3}12.4\% & \cellcolor[HTML]{F3FFF3}7.8\% & \cellcolor[HTML]{FFF4F3}17.5\% & \cellcolor[HTML]{F3FFF3}13.5\% & \cellcolor[HTML]{F3FFF3}13.6\% & \cellcolor[HTML]{F3FFF3}8.3\% & \cellcolor[HTML]{F3FFF3}23.0\% & \cellcolor[HTML]{F3FFF3}23.5\% & \cellcolor[HTML]{F3FFF3}21.4\% & \cellcolor[HTML]{F3FFF3}27.1\% & \cellcolor[HTML]{FFF4F3}6.6\% & \cellcolor[HTML]{FFF4F3}7.8\% & \cellcolor[HTML]{F3FFF3}4.4\% & \cellcolor[HTML]{FFF4F3}12.1\% \\
\textbf{CLT} & \cellcolor[HTML]{F3FFF3}0.0\% & \cellcolor[HTML]{FFF4F3}0.5\% & \cellcolor[HTML]{F3FFF3}0.0\% & \cellcolor[HTML]{FFF4F3}0.7\% & \cellcolor[HTML]{FFF4F3}0.0\% & \cellcolor[HTML]{F3FFF3}0.0\% & \cellcolor[HTML]{F3FFF3}0.0\% & \cellcolor[HTML]{F3FFF3}0.8\% & \cellcolor[HTML]{F3FFF3}1.4\% & \cellcolor[HTML]{F3FFF3}1.6\% & \cellcolor[HTML]{F3FFF3}1.4\% & \cellcolor[HTML]{F3FFF3}1.1\% & \cellcolor[HTML]{F3FFF3}0.5\% & \cellcolor[HTML]{F3FFF3}1.7\% & \cellcolor[HTML]{F3FFF3}1.1\% & \cellcolor[HTML]{FFF4F3}3.5\% \\
\textbf{CI} & 0.0\% & 0.0\% & 0.0\% & 0.0\% & \cellcolor[HTML]{FFF4F3}0.0\% & \cellcolor[HTML]{F3FFF3}0.0\% & \cellcolor[HTML]{F3FFF3}0.0\% & \cellcolor[HTML]{F3FFF3}0.0\% & \cellcolor[HTML]{F3FFF3}0.2\% & \cellcolor[HTML]{F3FFF3}0.0\% & \cellcolor[HTML]{F3FFF3}0.0\% & \cellcolor[HTML]{F3FFF3}1.1\% & 0.0\% & 0.0\% & 0.0\% & 0.0\% \\
\textbf{EM} & \cellcolor[HTML]{FFF4F3}8.2\% & \cellcolor[HTML]{F3FFF3}5.1\% & \cellcolor[HTML]{FFF4F3}24.8\% & \cellcolor[HTML]{F3FFF3}5.9\% & \cellcolor[HTML]{FFF4F3}7.7\% & \cellcolor[HTML]{FFF4F3}5.0\% & \cellcolor[HTML]{FFF4F3}21.6\% & \cellcolor[HTML]{FFF4F3}5.4\% & \cellcolor[HTML]{F3FFF3}1.4\% & \cellcolor[HTML]{F3FFF3}1.6\% & \cellcolor[HTML]{F3FFF3}2.9\% & \cellcolor[HTML]{F3FFF3}2.9\% & 0.0\% & 0.0\% & \cellcolor[HTML]{FFF4F3}1.1\% & \cellcolor[HTML]{FFF4F3}2.1\% \\
\textbf{EH} & \cellcolor[HTML]{F3FFF3}14.3\% & \cellcolor[HTML]{F3FFF3}19.5\% & \cellcolor[HTML]{F3FFF3}15.3\% & \cellcolor[HTML]{FFF4F3}24.5\% & \cellcolor[HTML]{F3FFF3}15.5\% & \cellcolor[HTML]{F3FFF3}18.5\% & \cellcolor[HTML]{F3FFF3}14.1\% & \cellcolor[HTML]{FFF4F3}25.7\% & \cellcolor[HTML]{F3FFF3}17.2\% & \cellcolor[HTML]{FFF4F3}22.5\% & \cellcolor[HTML]{FFF4F3}25.3\% & \cellcolor[HTML]{FFF4F3}21.5\% & \cellcolor[HTML]{F3FFF3}2.2\% & \cellcolor[HTML]{FFF4F3}3.3\% & \cellcolor[HTML]{FFF4F3}2.7\% & \cellcolor[HTML]{FFF4F3}5.0\% \\
\textbf{MG} & \cellcolor[HTML]{F3FFF3}2.0\% & \cellcolor[HTML]{F3FFF3}1.5\% & \cellcolor[HTML]{F3FFF3}1.0\% & \cellcolor[HTML]{F3FFF3}2.6\% & \cellcolor[HTML]{F3FFF3}1.0\% & \cellcolor[HTML]{F3FFF3}1.5\% & \cellcolor[HTML]{F3FFF3}1.5\% & \cellcolor[HTML]{F3FFF3}2.5\% & \cellcolor[HTML]{F3FFF3}2.2\% & \cellcolor[HTML]{F3FFF3}2.7\% & \cellcolor[HTML]{F3FFF3}2.4\% & \cellcolor[HTML]{F3FFF3}2.7\% & \cellcolor[HTML]{FFF4F3}1.6\% & \cellcolor[HTML]{FFF4F3}1.1\% & \cellcolor[HTML]{FFF4F3}1.1\% & \cellcolor[HTML]{FFF4F3}3.5\% \\
\textbf{RP} & \cellcolor[HTML]{F3FFF3}2.0\% & \cellcolor[HTML]{F3FFF3}2.1\% & \cellcolor[HTML]{F3FFF3}4.0\% & \cellcolor[HTML]{F3FFF3}3.0\% & \cellcolor[HTML]{F3FFF3}1.5\% & \cellcolor[HTML]{F3FFF3}2.5\% & \cellcolor[HTML]{F3FFF3}4.0\% & \cellcolor[HTML]{F3FFF3}2.9\% & \cellcolor[HTML]{F3FFF3}6.8\% & \cellcolor[HTML]{FFF4F3}16.5\% & \cellcolor[HTML]{FFF4F3}14.1\% & \cellcolor[HTML]{FFF4F3}10.7\% & 0.0\% & 0.0\% & 0.0\% & \cellcolor[HTML]{FFF4F3}0.7\% \\
\textbf{RA} & \cellcolor[HTML]{FFF4F3}1.0\% & \cellcolor[HTML]{F3FFF3}0.5\% & \cellcolor[HTML]{FFF4F3}1.0\% & \cellcolor[HTML]{FFF4F3}1.5\% & \cellcolor[HTML]{F3FFF3}0.5\% & \cellcolor[HTML]{F3FFF3}0.5\% & \cellcolor[HTML]{FFF4F3}1.0\% & \cellcolor[HTML]{FFF4F3}1.2\% & \cellcolor[HTML]{F3FFF3}0.0\% & \cellcolor[HTML]{F3FFF3}0.0\% & \cellcolor[HTML]{F3FFF3}0.0\% & \cellcolor[HTML]{F3FFF3}0.0\% & 0.0\% & 0.0\% & \cellcolor[HTML]{FFF4F3}0.5\% & \cellcolor[HTML]{FFF4F3}0.7\% \\
\textbf{SE} & \cellcolor[HTML]{FFF4F3}1.0\% & \cellcolor[HTML]{F3FFF3}0.0\% & \cellcolor[HTML]{FFF4F3}1.5\% & \cellcolor[HTML]{FFF4F3}1.5\% & \cellcolor[HTML]{F3FFF3}1.0\% & \cellcolor[HTML]{F3FFF3}0.5\% & \cellcolor[HTML]{F3FFF3}1.0\% & \cellcolor[HTML]{F3FFF3}1.2\% & \cellcolor[HTML]{F3FFF3}0.6\% & \cellcolor[HTML]{F3FFF3}0.2\% & \cellcolor[HTML]{F3FFF3}0.4\% & \cellcolor[HTML]{F3FFF3}0.9\% & \cellcolor[HTML]{FFF4F3}0.5\% & \cellcolor[HTML]{FFF4F3}0.6\% & \cellcolor[HTML]{FFF4F3}1.1\% & \cellcolor[HTML]{FFF4F3}2.1\% \\
\textbf{EA} & \cellcolor[HTML]{F3FFF3}16.8\% & \cellcolor[HTML]{F3FFF3}14.4\% & \cellcolor[HTML]{F3FFF3}11.4\% & \cellcolor[HTML]{F3FFF3}20.8\% & \cellcolor[HTML]{F3FFF3}17.0\% & \cellcolor[HTML]{F3FFF3}13.0\% & \cellcolor[HTML]{F3FFF3}11.6\% & \cellcolor[HTML]{F3FFF3}25.3\% & \cellcolor[HTML]{F3FFF3}24.6\% & \cellcolor[HTML]{F3FFF3}28.7\% & \cellcolor[HTML]{F3FFF3}20.8\% & \cellcolor[HTML]{F3FFF3}35.1\% & \cellcolor[HTML]{F3FFF3}7.7\% & \cellcolor[HTML]{F3FFF3}8.3\% & \cellcolor[HTML]{F3FFF3}6.6\% & \cellcolor[HTML]{FFF4F3}15.6\% \\
\textbf{LT} & \cellcolor[HTML]{F3FFF3}31.6\% & \cellcolor[HTML]{F3FFF3}44.1\% & \cellcolor[HTML]{F3FFF3}32.7\% & \cellcolor[HTML]{F3FFF3}55.8\% & \cellcolor[HTML]{F3FFF3}33.0\% & \cellcolor[HTML]{F3FFF3}46.0\% & \cellcolor[HTML]{F3FFF3}35.2\% & \cellcolor[HTML]{F3FFF3}57.7\% & \cellcolor[HTML]{F3FFF3}30.1\% & \cellcolor[HTML]{F3FFF3}26.0\% & \cellcolor[HTML]{F3FFF3}27.1\% & \cellcolor[HTML]{F3FFF3}32.4\% & \cellcolor[HTML]{F3FFF3}14.2\% & \cellcolor[HTML]{F3FFF3}16.7\% & \cellcolor[HTML]{F3FFF3}13.7\% & \cellcolor[HTML]{FFF4F3}22.0\% \\
\textbf{DA} & \cellcolor[HTML]{FFF4F3}6.1\% & \cellcolor[HTML]{FFF4F3}1.5\% & \cellcolor[HTML]{FFF4F3}1.5\% & \cellcolor[HTML]{FFF4F3}1.9\% & \cellcolor[HTML]{FFF4F3}5.2\% & \cellcolor[HTML]{FFF4F3}2.5\% & \cellcolor[HTML]{FFF4F3}2.0\% & \cellcolor[HTML]{FFF4F3}2.5\% & \cellcolor[HTML]{F3FFF3}6.4\% & \cellcolor[HTML]{F3FFF3}4.5\% & \cellcolor[HTML]{F3FFF3}5.1\% & \cellcolor[HTML]{F3FFF3}7.2\% & \cellcolor[HTML]{FFF4F3}2.2\% & \cellcolor[HTML]{FFF4F3}1.7\% & \cellcolor[HTML]{F3FFF3}0.5\% & \cellcolor[HTML]{FFF4F3}2.8\% \\
\textbf{UT} & \cellcolor[HTML]{F3FFF3}14.8\% & \cellcolor[HTML]{F3FFF3}12.3\% & \cellcolor[HTML]{FFF4F3}30.7\% & \cellcolor[HTML]{F3FFF3}17.8\% & \cellcolor[HTML]{FFF4F3}12.9\% & \cellcolor[HTML]{F3FFF3}10.5\% & \cellcolor[HTML]{FFF4F3}24.1\% & \cellcolor[HTML]{FFF4F3}16.6\% & \cellcolor[HTML]{F3FFF3}17.8\% & \cellcolor[HTML]{F3FFF3}16.7\% & \cellcolor[HTML]{F3FFF3}19.4\% & \cellcolor[HTML]{F3FFF3}20.6\% & 0.0\% & 0.0\% & \cellcolor[HTML]{FFF4F3}1.6\% & \cellcolor[HTML]{FFF4F3}2.1\% \\
\textbf{RO} & \cellcolor[HTML]{F3FFF3}1.5\% & \cellcolor[HTML]{F3FFF3}1.5\% & \cellcolor[HTML]{F3FFF3}2.0\% & \cellcolor[HTML]{F3FFF3}2.2\% & \cellcolor[HTML]{F3FFF3}1.0\% & \cellcolor[HTML]{F3FFF3}1.5\% & \cellcolor[HTML]{F3FFF3}2.5\% & \cellcolor[HTML]{F3FFF3}2.9\% & \cellcolor[HTML]{F3FFF3}3.6\% & \cellcolor[HTML]{F3FFF3}3.3\% & \cellcolor[HTML]{F3FFF3}3.7\% & \cellcolor[HTML]{F3FFF3}4.0\% & \cellcolor[HTML]{FFF4F3}1.6\% & \cellcolor[HTML]{FFF4F3}1.1\% & \cellcolor[HTML]{FFF4F3}1.1\% & \cellcolor[HTML]{FFF4F3}2.8\% \\
\textbf{MNT} & \cellcolor[HTML]{FFF4F3}98.5\% & \cellcolor[HTML]{FFF4F3}98.5\% & \cellcolor[HTML]{FFF4F3}98.0\% & \cellcolor[HTML]{F3FFF3}91.8\% & \cellcolor[HTML]{FFF4F3}97.9\% & \cellcolor[HTML]{FFF4F3}97.5\% & \cellcolor[HTML]{FFF4F3}98.5\% & \cellcolor[HTML]{F3FFF3}95.0\% & \cellcolor[HTML]{F3FFF3}91.2\% & \cellcolor[HTML]{FFF4F3}96.9\% & \cellcolor[HTML]{FFF4F3}99.0\% & \cellcolor[HTML]{FFF4F3}96.4\% & \cellcolor[HTML]{F3FFF3}18.6\% & \cellcolor[HTML]{F3FFF3}21.1\% & \cellcolor[HTML]{F3FFF3}18.0\% & \cellcolor[HTML]{FFF4F3}29.1\%
 \\ \bottomrule
\end{tabular}
\end{table}

 \section{Discussion}\label{sec:discussion}




\noindent-- \textbf{LLMs \textit{vs.} Evosuite}: Across all the studied dimensions, LLMs performed \textit{worse} than Evosuite. One reason is that LLMs do not always produce compilable unit tests (\tblref{tab:rq1-compile}). For example, while Evosuite produced one unit test for each of the \nJavaHumanEval{} classes under test, GPT-3.5-Turbo only produced \textbf{130} compilable (\ie executable) unit tests. Another reason is that LLMs do not seem to pay attention to the current MUT's implementation. A piece of evidence for this is that scenario 3 (which does not include the MUT's implementation) has better compilation rates than the rest. However, we also observed that GPT-3.5-Turbo generated test cases for ``stress-testing'', \eg using \codeJava{Integer.MAX_VALUE} to test for the MUT's behavior in the face of exceptionally large inputs.

\noindent-- \textbf{ Codex and StarCoder perform \textit{better} than \gpt}. This can be explained by the fact that  Codex and StarCoder are LLMs fine-tuned for code-related tasks in contrast to \gpt{}, which is tailored to dialogues (natural language). 

\noindent-- \textbf{LLMs  often ``hallucinate'' inexistent types, methods, \etc} For both datasets, the most common compilation error was due to missing symbols. For instance,  Codex generated inputs whose type were \codeJava{Tuple}, \codeJava{Pair}, \codeJava{Triple}, \codeJava{Quad}, and \codeJava{Quint}, which are non-existent in Java's built-in class types.


\noindent-- \textbf{Synergy between LLMs and TDD}. Although LLMs did not  achieve coverages or compilation rates comparable to Evosuite, they can still be useful as a starting point for TDD.  As we showed in our RQ2, LLMs can generate tests based on the MUT's JavaDoc. However, given the low correctness rates of LLMs, developers would still need to adjust the generated tests manually.

Given these findings, we observe a need for future research to focus on helping LLMs in reason over data types and path feasibility, as well as exploring the combination of SBST and LLMs for TDD. 
Furthermore, a recent study~\cite{albert22} surveyed 2,000 developers and analyzed anonymous user data, showing that GitHub Copilot makes developers more productive because the generated code can automate repetitive tasks. Thus, our findings provide some initial evidence that \textit{practitioners} following a TDD approach could benefit from LLM-generated tests as a means to
speed up their testing. Although further user studies would be needed to verify this hypothesis.

 \subsection{Threats to Validity}\label{sec:validity}

Creating canonical solutions for the Java samples in the HumanEval dataset~\cite{2023multilingual} introduced an internal validity threat. To mitigate it, we extensively vetted our solution with a test set provided by the dataset. Another validity threat relates to the use of  the SF110 benchmark \cite{evosuite}, JaCoCo \cite{jacoco} for calculating coverage results and TsDetect \cite{peruma2020tsdetect} to find test smells. In this case,  our analyses depend on the representativeness of the SF110 dataset (construct validity threat) and the accuracy of these tools.  However, the SF110 dataset is commonly used to benchmark automated test generation tools \cite{evosuite, bruce2019dorylus, shahabi2017performance} and JaCoCo and TsDetect are state-of-the-art tools~\cite{bilal2021jacoco,virginio2021test}.

 \section{Related Work}
\label{sec:related}

Previous works have focused on creating source code that can do a specific task automatically (code generation).
The deductive synthesis approach \cite{green69, manna2017}, in which the task specification is transformed into constraints, and the program is extracted after demonstrating the satisfaction of the constraints, is one of the foundations of program synthesis \cite{gulwani2017program}. Recurrent networks were used by Yin \etal \cite{yin2017} to map text to abstract syntax trees, which were subsequently coded using attention. A variety of large language learning models have been made public to generate code (e.g., CodeBert \cite{codebert}, CodeGen \cite{Nijkamp2022ACP} and CodeT5 \cite{codet5}) after being refined on enormous code datasets. Later, GitHub Copilot developed an improved auto-complete mechanism using the upgraded version of Codex \cite{chen2021codex}, which can help to solve fundamental algorithmic problems \cite{dakhel2022github}. 
Our work focuses not on code generation but on how a publicly available code generation tool can be used for specialized tasks like unit test generation without fine-tuning (\ie zero-shot test generation).

Shamshiri \etal\cite{sina18unittest} proposed a search-based approach that automatically generates tests that can reveal functionality changes, given two program versions. On the other hand, Tufano \etal~\cite{tufano@2020} proposed an approach that aims to generate unit test cases by learning from real-world focal methods and developer-written test cases. Pacheco \etal~\cite{pacheco@2007} presented a technique that improves random test generation by incorporating feedback obtained from executing test inputs as they are created for generating unit tests. Lu \etal \cite{lu23deepqtest} worked on testing autonomous driving systems with reinforcement learning.  
Lima \ et \cite{lima2023testing} surveyed the practitioners on software testing and refactoring.   In our work, we focus on zero-shot unit test generation using different contexts in order to measure the LLM's ability to generate compilable, correct, and smell-free tests. 

Sch{\"a}fer \etal \cite{schafer@2023} used Codex \cite{chen2021codex} to automatically generate unit tests using an adaptive approach. They used 25 npm packages to evaluate their tool, TESTPILOT. However, they evaluated their model only on statement coverage. They did not provide insight into the quality of the generated test cases and the choice of using a specific prompt structure. Lemieux \etal~\cite{lemieux@2023} combined the Search-based software testing (SBST) technique with the LLM approach. It explored whether Codex can be used to help SBST's exploration. Nashid \etal ~\cite{nashidretrieval@2023} aimed to devise an effective prompt to help large language models with different code-related tasks, \ie program repair and test assertion generation. Their approach provided examples of the same task and asked the LLM to generate code for similar tasks. Li \etal \cite{li2023nuances} used ChatGPT \cite{chatgpt} to find failure-inducing tests with differential prompting. Barei{\ss} \etal~\cite{bareiss@2022} performed a systematic study to evaluate how a pre-trained language model of code, Codex, works with code mutation, test oracle generation from natural language documentation, and  test case generation using few-shot prompting like Nashid \etal ~\cite{nashidretrieval@2023}. However, the benchmark has only 32 classes, so the findings may not be generalized. This work provides direction toward using examples of usage or similar tasks as a context. However, in a real case, there may not be any example of using the method and class that can be used in the prompt, and creating an example of a similar task needs human involvement. 
Our work focused on different contexts taken from the code base. We evaluated the quality of the generated unit tests not only on coverage and correctness but also based on the presence of test smells.

 \section{Conclusion}
\label{sec:conclusion}

We studied the capability of three code generation LLMs for  unit test generation. We conducted experiments with different contexts in the prompt and compared the results based on compilation rate, test correctness, coverage, and test smells. These models have a close performance with the state-of-the-art test generation tool for the HumanEval dataset, but their performance is poor for open-source projects from Evosuite based on coverage. Though our developed heuristics can improve the compilation rate, several generated tests were not compilable. Moreover, they heavily suffer from test smells like Assertion Roulette and Magic Number Test. In future work, we will explore how to enhance LLMs to understand language semantics better in order to increase test  correctness and compilation rates.

\bibliographystyle{ACM-Reference-Format}
\bibliography{reference}

\end{document}